\documentclass[11pt]{article}
\usepackage[a4paper]{geometry}
\usepackage[bf,SL,BF]{subfigure}
\usepackage{amsthm}
\usepackage{amsmath}
\usepackage{amssymb}
\usepackage{amsfonts}
\usepackage{graphicx}
\usepackage{graphics}

\def\eref#1{(\ref{#1})}
\def\hypoa{{\bf{I}}}
\def\hypob{{\bf{II}}}
\def\hypoc{{\bf{III}}}
\def\hypocf{{\bf{IV}}}

\def\N{\mathbb{N}}

\def\Pe{\bf{Pe}}

\def\P{\mathbb{P}}
\def\R{\mathbb{R}}
\def\E{\mathbb{E}}
\def\T{\mathbb{T}}
\def\Sp{\mathbb{S}}

\def\Z{\mathbb{Z}}
\def\M{\mathcal{M}}

\def\S{\mathcal{S}}
\def\SL{\mathcal{S}_{L^\infty}}

\def\<{\big\langle}
\def\>{\big\rangle}

\def\diiv{\operatorname{div}}

\def\sym{{\operatorname{sym}}}

\def\Vol{{\operatorname{Vol}}}

\newtheorem{Theorem}{Theorem}[section]

\newtheorem{Lemma}[Theorem]{Lemma}
\newtheorem{Corollary}[Theorem]{Corollary}

\newtheorem{Proposition}[Theorem]{Proposition}

\theoremstyle{remark}
\newtheorem{Remark}[Theorem]{Remark}
\theoremstyle{definition}

\newtheorem{Definition}[Theorem]{Definition}

\newcommand{\figbox}[1]{%
  \fbox{%
    \vbox to 1in{%
    \vfil
    \hbox to 2in{%
      \hfil
      #1%
      \hfil}%
    \vfil}}}

\newcommand{\goodgap}{%
  \hspace{\subfigcapskip}}

\begin{document}

\title{Averaging versus Chaos in  Turbulent Transport?\protect\footnotetext{AMS 1991 {\it{Subject Classification}}.
 Primary 76F25; secondary   76F30, 76F20, 35B27} \protect\footnotetext{{\it{Key words and phrases}}.
  Turbulence, passive transport, super-diffusion, multi-scale homogenization, renormalization,
   chaos, turbulent diffusivity, eddy viscosity, anisotropic turbulence, tornado.}}         
\author{Houman Owhadi\footnote{
LATP, UMR CNRS 6632, CMI, Universit\'{e} de Provence , owhadi@cmi.univ-mrs.fr }}        
\date{\today}          
\maketitle \abstract{ In this paper we  analyze the transport of  passive tracers by deterministic stationary
incompressible flows which can be decomposed over an infinite number of spatial scales without separation
between them. It appears that a low order dynamical system related to local Peclet numbers can be extracted from
these flows and it controls their transport properties. Its analysis shows that these flows are strongly
self-averaging and super-diffusive: the delay $\tau(r)$ for any finite number of passive tracers initially close
to separate till a distance $r$ is almost surely anomalously fast ($\tau(r)\sim r^{2-\nu}$, with $\nu>0$). This
strong self-averaging property is such that the dissipative power of the flow compensates its convective power
at every scale. However as the circulation increase in the eddies the transport behavior of the flow may
(discontinuously) bifurcate and become ruled by deterministic chaos: the self-averaging property
 collapses and  advection dominates dissipation. When the flow is anisotropic a new
formula describing turbulent conductivity is identified.}

\section{Introduction}
In this paper we study the passive transport in $\R^d$ ($d\geq 2$) of a scalar $T$ by a divergence free steady
vector field $v$ characterized by the following partial differential equation. ($\kappa>0$ being the molecular
conductivity)
\begin{equation}\label{dskjddjjccfdgg8121}
\partial_t T+v \nabla T= \kappa \Delta T
\end{equation}
 We will assume $v$ to be given by an infinite (or large) numbers of spatial scales without  any assumption
  of self-similarity \cite{Ave96}. It will be shown that one can
extract from the flow a low order dynamical system related to  local Peclet tensors which controls the transport
properties of the flow. Based on the analysis of this dynamical system we will show that the transport is almost
surely super-diffusive, that
 is to say, the time of separation of any finite number of passive tracers driven by the same flow and independent
 thermal noise behave like $r^{2-\nu}$ with $\nu>0$. Similar programs of investigations have shown that the mean squared
  displacement of a
 single particle is anomalously fast when averaged with respect to space, time and the randomness of the flow
 (\cite{MR98i:76046}, \cite{KoOl02}, \cite{Fa02}).
 The point here is to show that the transport is strongly
 self-averaging: the diffusive properties are anomalously fast (before being averaged with respect to the thermal
 noise, or a probability law of the flow), moreover the pair separation is also anomalously fast. The fast
 behavior of the transport of a single particle can be created by long distance correlations in the structure
 of the  velocity field but this is not sufficient to produce fast pair separation. In this paper
non-asymptotic estimates will be given, showing that the transport is
 controlled by a never-ending averaging phenomenon (\cite{Owh01}, \cite{Ow00a}, \cite{BeOw00b}, \cite{BeOw00c}).
 The analysis of the low order dynamical system allows to obtain a formula linking the minimal and maximal
  eigenvalues of the turbulent eddy diffusivity.  It will be shown that the transport properties
 depend only on the power law in $v$ and not on its particular geometry (which is not a priori obvious since we consider
  a quenched model).
However, depending on the geometrical characteristics of the eddies at each scale, as the flow rate
  is increased in these eddies we observe that the super-diffusive behavior may bifurcates
  towards a Chaotic transport: the multi-scale averaging picture collapses and the flow becomes highly unstable,
  sensitive
  to the characteristics of the microstructure  and dominated by convective terms.

\section{The Model}\label{qqqztfxsdstffatf1}
We want to analyze the properties of the solutions of  the following stochastic differential equation which is the Lagrangian formulation of the passive transport equation \eref{dskjddjjccfdgg8121}.
\begin{equation}\label{shsshkjjb81}
dy_t=\sqrt{2 \kappa} d\omega_t+\nabla. \Gamma(y_t)\,dt.
\end{equation}
Here $\kappa>0$ is the molecular conductivity of the flow, $\omega_t$ a standard Brownian Motion on $\R^d$
related to the thermal noise, $\Gamma$ is a skew-symmetric matrix on $\R^d$ called stream matrix of the flow and
$\nabla .\Gamma$ its divergence. Thus $\nabla. \Gamma$ is the divergence free drift defined by $(\nabla
.\Gamma)_i=\sum_{j=1}^d\partial_j \Gamma_{ij}$. We assume that $\Gamma$ is given by an infinite sum of periodic
stream matrices with (geometrically) increasing periods and increasing amplitude.
\begin{equation}\label{Modsubfracuinfty}
\Gamma=\sum_{k=0}^\infty \gamma_k E^{k}(\frac{x}{R_k}).
\end{equation}
In the formula \eref{Modsubfracuinfty} we have three important ingredients: the stream matrices $E^k$ (also
called eddies), the scale parameters $R_k$ and the amplitude parameters $\gamma_k$ (the stream matrices $E^k$
are dimension-less and the parameters $\gamma_k$ have the dimension of a conductivity). We will now describe the
hypothesis we make on these three items of the model. Let us write  $T^d:=\R^d/\Z^d$  the torus of dimension $d$
and side one and for $\alpha \in[0,1]$, $\S^\alpha(T^d)$  the space of $d\times d$ skew-symmetric matrices on
$T^d$ with $\alpha$-Holder continuous coefficients and $\|.\|_\alpha$ the norm associated to that space. For
$E\in \S^\alpha(T^d)$
\begin{equation}\label{ModsubddsContUngradUn}
\|E\|_\alpha:=\sup_{i,j \in \{1,\ldots,d\}}\sup_{x\not=y}|E_{ij}(x)-E_{ij}(y)|/|x-y|^\alpha.
\end{equation}
\renewcommand{\labelenumi}{\bf{\Roman{enumi}}}
\begin{enumerate}
\item\label{hypo1} \underline{Hypotheses on the stream matrices $E^k$}\\
There exists $0<\alpha\leq 1$ such that for all $k\in \N$
\begin{equation}\label{jhshdddikuou1}
E^k \in \S^\alpha(\T^d).
\end{equation}
The $\S^\alpha$-norm of the $E^k$ are uniformly bounded, i.e.
\begin{equation}\label{ModsubContUngradUn}
K_\alpha:=\sup_{k\in \N} \|E^k\|_\alpha <\infty.
\end{equation}
Moreover for all $k$,
\begin{equation}\label{jhshdikuou1}
E^k(0)=0.
\end{equation}
Observe that the $\S^0$-norms of the $E^k$ are uniformly bounded and we will write
\begin{equation}\label{ModsubContUngradUnGer}
K_0:=\sup_{k\in \N} \sup_{i,j \in \{1,\ldots,d\}} \|E^k\|_0
\end{equation}

\item\label{hypo2} \underline{Hypotheses on the scale parameters $R_k$}\\
$R_k$ is a spatial scale parameter growing exponentially fast with $k$,
more precisely we will assume that $R_0=r_0=1$ and that the ratios
between scales defined by
\begin{equation}\label{jahgsvagvjh6761}
 r_k= R_k/R_{k-1}\in \R^{*}
\end{equation}
for $k\geq 1$, are reals uniformly bounded away from $1$ and
$\infty$: we will denote by
\begin{equation}\label{Modsubboundrnrhonmin}
 \rho_{\min}:=\inf_{k\in \N^*} r_k \quad\text{and}\quad
\rho_{\max}:=\sup_{k\in \N^*} r_k
\end{equation}
and assume that
\begin{equation}\label{Modsubboud32ndrnrhonmin}
 \rho_{\min}\geq 2 \quad\text{and}\quad \rho_{\max}<\infty.
\end{equation}
\item\label{hypo3} \underline{Hypotheses on the flow rates $\gamma_k$}\\
$\gamma_k$ is an amplitude parameter (related to the local rate of the flow) growing exponentially fast with the scale $k$,
more precisely we will assume that $\gamma_0=1$ and that their ratios $\gamma_k/\gamma_{k-1}$
for $k\geq 1$, are positive reals uniformly bounded away from $1$ and
$\infty$: we will denote by
\begin{equation}\label{Modsubboundrnrhonmigamman}
 \gamma_{\min}:=\inf_{k\in \N^*} (\gamma_k/\gamma_{k-1}) \quad\text{and}\quad
\gamma_{\max}:=\sup_{k\in \N^*} (\gamma_k/\gamma_{k-1})
\end{equation}
and assume that
\begin{equation}\label{Modsubboud32ndrnrhonmigamman}
 \gamma_{\min}> 1 \quad\text{and}\quad \gamma_{\max}<\rho_{\min}^\alpha.
\end{equation}
\end{enumerate}
\renewcommand{\labelenumi}{\arabic{enumi}}

\begin{Remark}
The  uniform $\alpha$-Holder continuity of the stream matrices $E^k$ is sufficient to obtain a well defined
$\alpha$-Holder continuous stream matrix $\Gamma$, however $\Gamma$ is not differentiable in general. In this
case the stochastic differential equation \eref{shsshkjjb81} is formal. For the simplicity of the presentation
and to start with, when referring to the SDE \eref{shsshkjjb81} we will assume that
\begin{equation}\label{sajhskz71}
\alpha=1 \quad \text{and that the stream matrices $E^k$ are uniformly $C^1$.}
\end{equation}
It follows from the hypothesis \hypoa, \hypob and \hypoc \   \ that $\Gamma$ is a well defined uniformly $C^1$
skew-symmetric matrix on $\R^d$, thus the Stochastic Differential Equation \ref{shsshkjjb81}
is well defined and admits a unique solution.\\
 The differentiability hypothesis \eref{sajhskz71} though convenient in order to define the process $y_t$
  is in fact useless, the theorems are also meaningful and true for $0<\alpha<1$
 (since they will refer to the diffusion associated to the weakly defined operator $\nabla .(\kappa+\Gamma)\nabla$).
\end{Remark}

\begin{Remark}
Observe that  the power law of the flow in this paper is not Kolmogorov. Indeed if $v(l)$ is the velocity of the
eddies of size $l$ and $\mathcal{E}(k)$ the kinetic energy distribution in the Fourier modes then with the
Kolmogorov law one should have
$$v(l)\sim l^\frac{1}{3}\quad \text{and} \quad \mathcal{E}(k)\sim k^{-\frac{5}{3}}.$$
In our Model we have
$$v(l) \sim l^{\frac{\ln \gamma}{\ln \rho}-1}\quad \quad \mathcal{E}(k)\sim k^{1-2 \frac{\ln \gamma}{\ln \rho} }.$$
Thus to be consistent with a Kolmogorov spectrum one should have $\gamma=\rho^\frac{4}{3}$ this case will be
analyzed in a forthcoming paper.
\end{Remark}

\begin{figure}[htbp]
\centerline{\includegraphics[width=0.4\textwidth,height= 0.4\textwidth]{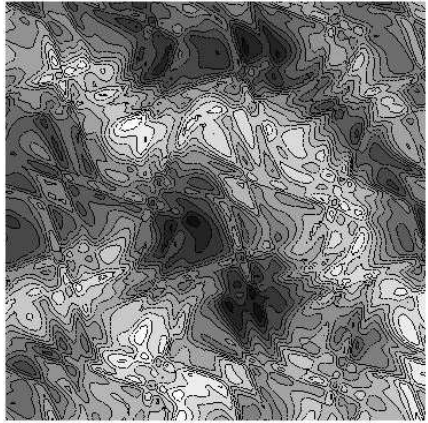}} \caption{A simple example
of the multiscale flow \label{particullll}}
\end{figure}

As an example, we have illustrated in the figure \ref{particullll}
the contour lines of a two scale flow with stream function
$h_0^2(x,y)=\sum_{k=0}^2 \gamma^k
h(\frac{x}{\rho^k},\frac{y}{\rho^k})$ , with $\rho=3$,
$\gamma=1.1$ and $h(x,y)=2\sin(2\pi x+3\cos(2\pi y-3\sin(2\pi
x+1)))\sin(2\pi y+3\cos(2\pi x-3\sin(2\pi y+1)))$

\section{A reminder on the eddy conductivity}\label{djgddffdgdd7766121}
We write $\M_{d,\sym}$ the space of $d\times d$ symmetric elliptic constant matrices and $\SL(\T^d_R)$ the space of skew-symmetric matrices  with coefficients in $L^\infty(\T^d_R)$ ($\T^d_R:=R \T^d$ stands for the torus of dimension d and side $R$).   For $a\in \M_{d,\sym}$ and $E$ a skew symmetric matrix with bounded coefficients the heat kernel associated to the passive transport operator $\nabla .(a+E)\nabla$ (defined in a weak sense) is Gaussian by Aronson estimates \cite{Nor97}. We will now assume $E$ to be periodic: $E\in \SL(\T^d_R)$. In this case the process associated to the operator $L=\nabla .(a+E)\nabla$ exhibits self-averaging properties and  we will note  $\sigma_{\sym}(a,E)$ the effective conductivity  associated to the homogenization of that operator   \cite{BeLiPa78}, \cite{JiKoOl91}. Writing $p(t,x,y)$ the heat kernel associated to $L$, it is well known  that  $\sigma_{\sym}(a,E)$ is a $d\times d$ elliptic symmetric matrix satisfying, for all $x,l\in \R^d$
\begin{equation}
|l|^2_{\sigma_{\sym}(a,E)}=\frac{1}{2}\lim_{t\rightarrow \infty}t^{-1}\int_{\R^d}p(t,x,y)(y.l)^2 \,dy.
\end{equation}
We have used the notation $|l|_a^2:={^tl al}$. If $z_t$ is the process generated by $L$, then as $\epsilon
\downarrow 0$, $\epsilon z_{t/{\epsilon^2}}$ converges in law to a Brownian motion with covariance matrix
$D(a,E)$ called effective diffusivity and proportional to the effective conductivity.
\begin{equation}
D(a,E)=2 \sigma_{\sym}(a,E).
\end{equation}
Let us remind that  $\sigma_{\sym}(a,E)$  is given by the following variational formula (\cite{Nor97} lemma
3.1): for $\xi \in \R^d$
\begin{equation}\label{HoCoeDNorPre12}
|\xi|_{\sigma_{\sym}^{-1}(a,E)}^2=\inf_{(f,H)\in C^\infty(\T^d_R)\times \S(\T^d_R)}
R^{-d}\int_{\T^d_R}|\xi-\nabla .H+(a+E)\nabla f|_{a^{-1}}^2\,dx.
\end{equation}
Where we have written $\S(\T^d_R)$ the space of skew symmetric matrices with coefficients in $C^\infty(\T^d_R)$.\\
The symmetric tensor $\sigma_{\sym}(a,E)$ is also called eddy conductivity: after averaging the information on
particular geometry of the eddies associated to $E$ is lost and the conductivity of the flow is replaced by an
increased conductivity  $\sigma_{\sym}(a,E)$. Let us define for $P \in \SL(\T^d)$ and $\rho \in \R^*$, $S_\rho P
\in \SL(\T^d_\frac{1}{\rho})$ by
\begin{equation}\label{ddhdsvvdg87efeeff}
S_\rho P(x):=P(\rho x).
\end{equation}
 It is important to note that the effective conductivity is invariant by scaling, i.e. $\sigma_{\sym}(a,S_\rho P)=\sigma_{\sym}(a, P)$; thus we can assume for simplicity that $R=1$ and $E\in \SL(\T^d)$.
When $E$ is smooth $\sigma_{\sym}(a,E)$ is given \cite{BeLiPa78} by solving the following cell problem:
\begin{equation}
\nabla .(a+E)(l-\nabla \chi_l^{a,E})=0.
\end{equation}
Where $l\in \R^d$, $\chi_l^{a,E}\in C^\infty(\T^d)$ and $\int_{\T^d} \chi_l^{a,E}(x)\,dx=0$. Write
$F_l^{a,E}=l.x-\chi_l^{a,E}(x)$, observe that $F_l^{a,E}$ is linear in $l$, thus we will write $F^{a,E}$ the
vector field $(F^{a,E})_i:=F_{e_i}^{a,E}$ and $\nabla F^{a,E}$ the matrix $(\nabla F^{a,E})_{ij}:=\partial_i
F^{a,E}_{e_j}$. The eddy conductivity is then given by
\begin{equation}\label{dskjsdbjdb89018}
\sigma_{\sym}(a,E)=\int_{\T^d}{^t\nabla F^{a,E}(x)a\nabla F^{a,E}(x)}\,dx.
\end{equation}
Let us remind that the matrix $\sigma(a,E)$ defined by
\begin{equation}
\sigma(a,E):=\int_{\T^d}(a+E(x))\nabla F^{a,E}(x)\,dx
\end{equation}
is called the flow effective conductivity \cite{FaPa94} and is also given by the following variational formula \cite{Nor97}:
 for $\xi,l \in \R^d$
\begin{equation}\label{dskbosdinbopin91}
\begin{split}
|\xi-\sigma(a,E) l|_{\sigma_{sym}^{-1}(a,E)}^2:=\inf_{(f, H)\in C^\infty(\T^d)\times
\S(\T^d)}\int_{\T^d}|\xi-\nabla .H-(a+E)(l-\nabla f)|_{a^{-1}}^2\,dx.
\end{split}
\end{equation}
It is easy to check that  $\sigma_{\sym}(a,E)$ is the symmetric part of $\sigma(a,E)$ which implies the
following variational formulation for the eddy conductivity
\begin{equation}\label{HoCoeDNorPre14}
|l|_{\sigma_{\sym}}^2=\inf_{\xi \perp l, (f,H) \in C^\infty(\T^d)\times \S(\T^d)}\int_{\T^d}|\xi-\nabla
.H-(a+E)(l-\nabla f)|^2_{a^{-1}}\,dx.
\end{equation}
Where we have written $\xi\perp l$ is the subspace of $\xi \in \R^d$ orthogonal to the vector $l$: $\xi.l=0$.

\section{Main results}\label{dkjdhdcuiuhc7c787121}
\subsection{Averaging with two scales}\label{eexheuzxxezugxzegx}
 Let
$a\in \M_{d,\sym}$, $P\in \SL(\T^d)$ and $K\in \S^\alpha(\T^d)$.  We will prove in subsection
\ref{skndkndkdndn8981111} the following estimate of $\sigma_{\sym}(a,S_R P+ K)$ the effective conductivity for a
two-scale medium when $R$ is an integer (and $S_R$ is the scaling operator \eref{ddhdsvvdg87efeeff}).
\begin{Theorem}\label{0kjhhbxdhbo1222}
There exists a function $f:\R^2\rightarrow \R^+$ increasing in each of its arguments such that for $a\in
\M_{d,\sym}$, $R\in \N^*$, $P\in \SL(\T^d)$ and $K\in \S^\alpha(\T^d)$
\begin{equation}\label{dkjdjbj8981111}
\begin{split}
\left(1+\epsilon\left(R\right)\right)^{-4} \sigma_{\sym}\big(\sigma_{\sym}(a,P), K\big) \leq \sigma_{\sym}(a,S_R
P+ K) \leq
 \sigma_{\sym}\big(\sigma_{\sym}(a,P), K\big) (1+\epsilon(R))^4.
\end{split}
\end{equation}
\begin{equation}\label{sdjddhdgdgz7112}
\begin{split}
\text{With} \quad \epsilon(R)= \big(\frac{\|K\|_\alpha}{R^\alpha \lambda_{\min}(a) }\big)^\frac{1}{2}
f(d,\frac{\|a+P\|_\infty}{\lambda_{\min}(a)}).
\end{split}
\end{equation}
\end{Theorem}
\begin{Remark}
Theorem \ref{0kjhhbxdhbo1222} implies obviously that
\begin{equation}\label{dkjdjbj8981111000}
\begin{split}
\sigma_{\sym}\big(\sigma_{\sym}(a,P), K\big)= \lim_{R\rightarrow \infty} \sigma_{\sym}(a,S_R P+ K).
\end{split}
\end{equation}
Thus $\sigma_{\sym}\big(\sigma_{\sym}(a,P), K\big)$ should be interpreted as the effective conductivity of the two-scale flow with a complete separation of scales. So we will also write it $\sigma_{\sym}(a,S_\infty P+ K)$. Naturally $\sigma_{\sym}\big(\sigma_{\sym}(a,P), K\big)$ is also computable from an explicit cell problem (see \eref{dskjsdbjdb89018}).
\end{Remark}

\begin{figure}[htbp]%
  \begin{center}%
    \subfigure[stream lines of $S_R P$ and $K$.\label{fig:AfirstA}]{\includegraphics[width=0.38\textwidth,height= 0.38\textwidth]{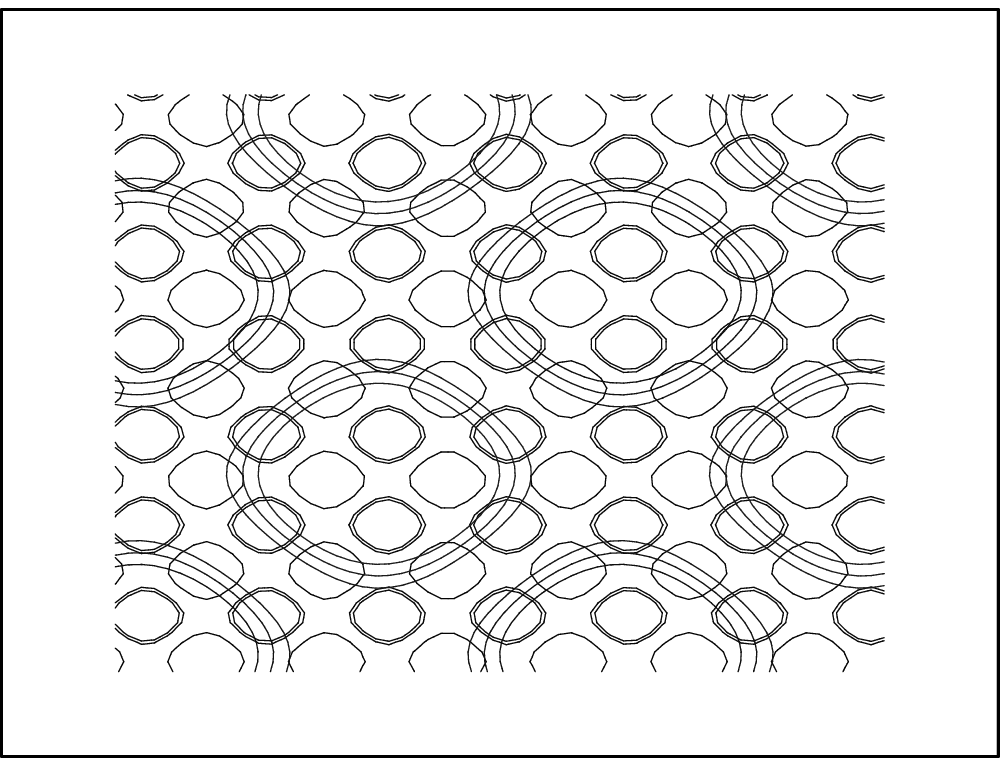}}%
    \goodgap
    \subfigure[stream lines of $S_R \Theta_y P$ and $K$.\label{fig:AfirstB}]{\includegraphics[width=0.38\textwidth,height= 0.38\textwidth]{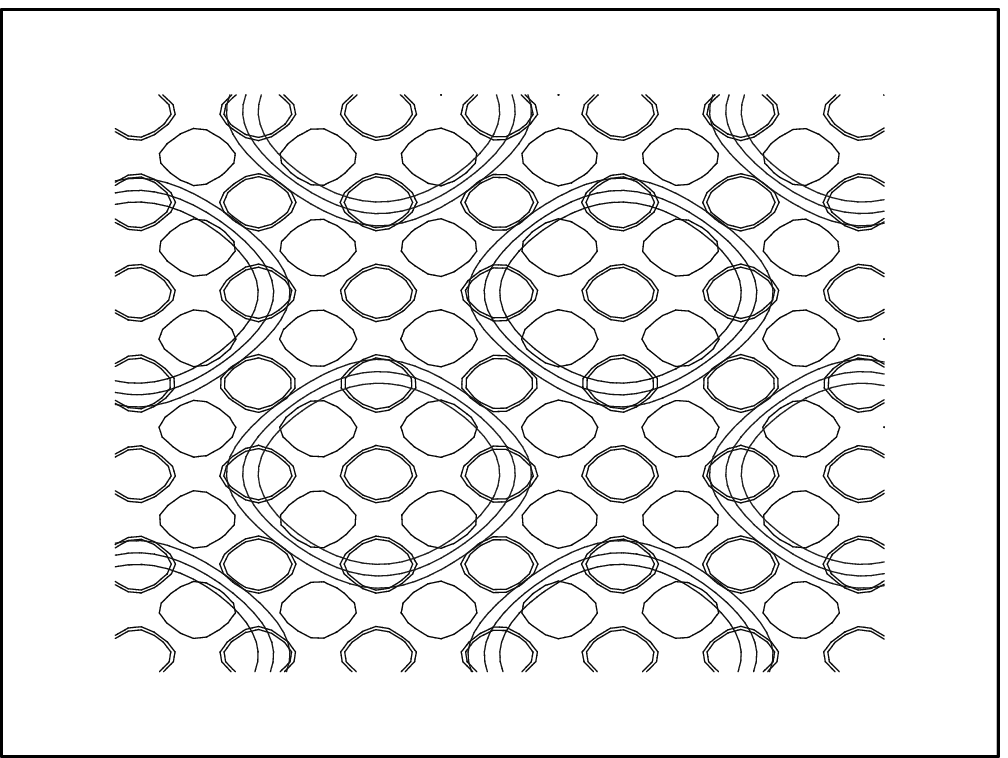}}\\[-10pt]
    \caption{Two scales flow.}%
    \label{fig:Afirst}%
\end{center}
\end{figure}

\paragraph{Averaging versus chaotic coupling}
The equation \eref{dkjdjbj8981111} basically says that when $\epsilon(R)$ is small, the mixing length of the
smaller scale $P(Rx)$ is smaller than scale at which the fluctuations of the larger scale $K(x)$ start to be
felt.
 Now it is very important to observe that as $\lambda_{\min}(a)\downarrow 0$, $\epsilon(R)$ explode towards
 infinity and this collapse of the two-scale averaging is not an artefact, it is easy to see that the
  estimate \eref{sdjddhdgdgz7112} is sharp. What happens is a transition from averaging to a
  chaotic coupling between the two scales.
  More precisely as $\lambda_{\min}(a)\downarrow 0$, the mixing length of the smaller scale explode
   well above the visibility length of the larger scale, the two scales are no longer separated in
   the averaging and their particular geometry  can no longer be ignored (collapse of the averaging paradigm).
    Moreover writing for $y\in [0,1]^d$, $\Theta_y$ the translation operator acting on functions
    $f$ of $\R^d$ by $\Theta_y f(x)= f(x+y)$, observe that in the
    limit of complete separation between scales the two-scale averaging is invariant
    with respect to a relative translation of one scale with respect to an other:

\begin{equation}\label{dkjdjbj89ss81111000}
\begin{split}
\lim_{R\rightarrow \infty} \sigma_{\sym}(a,S_R \Theta_y P+ K)=\lim_{R\rightarrow \infty} \sigma_{\sym}(a,S_R P+ K).
\end{split}
\end{equation}
But the limit $\lambda_{\min}(a)\downarrow 0$ is singular and this invariance by translation is lost: for $l\in \R^d$
\begin{equation}\label{dkjdjbj89sssdds81111000}
\begin{split}
\Big( {^tl}\sigma_{\sym}(a,S_R \Theta_y P+ K)l- {^tl}\sigma_{\sym}(a,S_R P+ K)l  \Big)\Big( {^tl}\sigma_{\sym}(a,S_R P+ K)l  \Big)^{-1}.
\end{split}
\end{equation}
may explode towards infinity. Indeed it is easy to see that for any $R\in \N^*$, there exist $P,K \in
\S^1(\T^d)$ with $\|P\|_1\leq C_d$, $\|K_1\|\leq C_d$  such that there exists $y\in [0,1]^d$ and $l\in \R^d$
with
\begin{equation}\label{dkjdjbj89ssss81111000}
\begin{split}
\lim_{\zeta\downarrow 0}\Big( {^tl}\sigma_{\sym}(\zeta I_d,S_R \Theta_y P+ K)l \Big)\Big( {^tl}\sigma_{\sym}(a,S_R P+ K)l  \Big)^{-1}=\infty.
\end{split}
\end{equation}
We have illustrated this symmetry breaking in the figures \ref{fig:Afirst} representing a two scale flow. In
figure \ref{fig:AfirstA} the larger eddies are surrounded by a non void region where the flow is null and
asymptotic behavior of the effective conductivity at vanishing molecular conductivity  is given by
\begin{equation}\label{dkjdjbj89ssss81sss111000}
\begin{split}
\sigma_{\sym}(\zeta I_d,S_R  P+ K)\sim C_1 \zeta I_d.
\end{split}
\end{equation}
In figure \ref{fig:AfirstB} we have operated a small translation of the smaller scale with respect to the larger
one. The result of this relative translation is the percolation of stream lines of the flow:
 a particle driven
by the flow can go to infinity by following them. It follows  after this small perturbation that asymptotic
behavior of the effective conductivity of the two scale flow at vanishing molecular conductivity is given by

\eref{dkjdjbj89sssxxxxw81111000}.
\begin{equation}\label{dkjdjbj89sssxxxxw81111000}
\begin{split}
\sigma_{\sym}(\zeta I_d,S_R \Theta_y P+ K)\sim C_2 \zeta^\frac{1}{2} I_d.
\end{split}
\end{equation}
We call this sensibility with respect to the relative translation $\Theta_y$, \emph{chaotic coupling between scales}.
The asymptotic \eref{dkjdjbj89sssxxxxw81111000} can be understood from a boundary layer analysis (see
\cite{Chil79} and \cite{FaPa94}).

\subsection{Multiscale eddy conductivity and the \emph{renormalization core}.}
Let us write
\begin{equation}\label{Modsubfracuinfty222}
\Gamma^{0,n}=\sum_{k=0}^n \gamma_k E^{k}(\frac{x}{R_k}).
\end{equation}
For this subsection we will use the following hypothesis

\paragraph{IV}\underline{Hypothesis on the ratios between scales}: for all $k\in \N$, $r_k\in \N^*$.

\paragraph{}
Our objective is to obtain quantitative estimates the multi-scale eddy viscosities  $\big(\sigma(\Gamma^{0,n})\big)_{n\in \N}$; observe that under the hypothesis \hypocf, $\Gamma^{0,n}$ is periodic, thus its effective conductivity is well defined by equation \eref{HoCoeDNorPre12} (that is its only utility, we will not need this hypothesis to prove super-diffusion). These estimates (theorem \ref{asjhsskbjs801}) will be proven by induction on the number of scales. The basic step in this induction is the estimate \eref{dkjdjbj8981111} on the effective conductivity for a two scale periodic medium.
We will need to introduce a dynamical system called the \emph{renormalization core} which will play a central role in the  transport properties of the stochastic differential equation \eref{shsshkjjb81}.

\begin{Definition}\label{sdjhdsgdsuzdggggg2876261212}
We propose to call \emph{"renormalization core"} the dynamical system $(A^n)_{n\in \N}$  of $d\times d$ symmetric strictly elliptic matrices defined by

\begin{equation}\label{wdkbkwjdebkljb81}
A^0=\frac{\kappa}{\gamma_0} I_d \quad \text{and}\quad A^{n+1}=\frac{\gamma_{n}}{\gamma_{n+1}}\sigma_{sym}(A^{n},E^n).
\end{equation}
\end{Definition}
For $B$ a $d\times d$ symmetric coercive matrix let us define the function $g(B)$ by
\begin{equation}\label{hjdgffgfz63112}
g(B):=\big(\frac{K_\alpha}{ \lambda_{\min}(B) (1-\gamma_{\max}/\rho_{\min}^\alpha)}\big)^\frac{1}{2}f\big(d,(\lambda_{\max}(B)+K_0)/\lambda_{\min}(B)\big).
\end{equation}
Where $f$ is the function appearing in theorem \ref{0kjhhbxdhbo1222}.
We will prove in the subsection \ref{sksjkjbkjcbc987191} the following theorem.
\begin{Theorem}\label{asjhsskbjs801}
Under hypotheses \hypoa, \hypob, \hypoc and \hypocf  for all $n\in \N^*$
\begin{equation}
\begin{split}
\gamma_{n+1} A^{n+1} \prod_{p=1}^{n}(1+\epsilon_p)^{-4} \leq \sigma_{sym}(\kappa I_d,\Gamma^{0,n})\leq \gamma_{n+1} A^{n+1} \prod_{p=1}^{n}(1+\epsilon_p)^{4}
\end{split}
\end{equation}
with
\begin{equation}
\epsilon_p=\Big(\frac{\gamma_p}{\gamma_{p-1}r_p^{\alpha}}\Big)^\frac{1}{2} g(A^{p-1}).
\end{equation}
 $A^n$ being the \emph{renormalization core} \eref{wdkbkwjdebkljb81}
\end{Theorem}

Observe that $\gamma_{n+1} A^{n+1}$ is the etimate given by reiterated homogenization under the assumption of
complete separation between scales, i.e. $\rho_{\min}\rightarrow \infty$ and the error term
$\prod_{p=1}^{n}(1+\epsilon_p)^{4}$  controlled by the \emph{renormalization core} $A^{k}$ which reflects the
interaction
 between the scales $k$ and $k+1$. As $\lambda_{\min}(A_k)\downarrow 0$ one passes from a
 separation of the scales $k$ and $k+1$ to a chaotic coupling between this two scales.
 Moreover it is easy to obtain from theorem \ref{asjhsskbjs801} that
\begin{equation}\label{sdhdjhdgd7777332}
\lim_{r_1,\ldots,r_{n-2} \rightarrow \infty} \sigma_{sym}(\kappa I_d,\Gamma^{0,n})= \gamma_{n-1} \sigma_{sym}(A^{n-1},S_{r_n} E^{n-1} +\frac{\gamma_n}{\gamma_{n-1}} E^n).
\end{equation}
Assume that the multi-scale averaging scenario holds and $\rho_{\min}<\infty$. In that scenario,
 $\sigma_{sym}(\kappa I_d,\Gamma^{0,n})$ can be approximated by its limit at asymptotic
 separation between scales. We obtain a contradiction if $\liminf_{n\rightarrow \infty} \lambda_{\min}(A^n)=0$
 from \eref{sdhdjhdgd7777332} and the collapse of the two-scale averaging scenario  given in
  subsection \ref{eexheuzxxezugxzegx} and figure \ref{fig:Afirst}. In other words
   if $\liminf_{n\rightarrow \infty} \lambda_{\min}(A^n)=0$ then the self-averaging property of the flow
   collapses towards a chaotic coupling between all the scales which is characterized by the breaking of the
   invariance by relative translation between the scales.

\subsubsection{What is the \emph{renormalization core}?}
 First observe that it
is a dimensionless tensor. At the limit of infinite separation between scales the eddy conductivity created by
the scales $0,\ldots,n-1$ is $\lim_{\rho_{\min}\rightarrow \infty}\sigma_{\sym}(\kappa I_d, \Gamma^{0,n-1})$.
The typical scale length associated to the scale $n$ is $R_n$ and the velocity of the flow at this scale is of
the order of $\gamma_n R_n^{-1}$ (we assume $K_1$ to be of order one). Thus  at the scale $n$ one can define a
local renormalized Peclet tensor $\Pe^n$ by:
\begin{equation}
\Pe^n:=R_n\times \frac{\gamma_n}{R_n}\times \big(\lim_{\rho_{\min}\rightarrow \infty}\sigma_{\sym}(\kappa I_d,
\Gamma^{0,n-1})\big)^{-1}.
\end{equation}
But at the limit of complete separation between scales $(A^n)^{-1}$ is equal to the ratio between the convective
strength $\gamma_n$ of the scale $n$ and the local turbulent conductivity at the scale $n-1$:
\begin{equation}
(A^n)^{-1}= \lim_{\rho_{\min}\rightarrow \infty}\gamma_{n} \big(\sigma_{\sym}(\kappa I_d, \Gamma^{0,n-1})\big)^{-1}.
\end{equation}
It follows that
\begin{equation}
(A^n)^{-1}= \Pe^n.
\end{equation}
Thus one can interpret  the \emph{renormalization core} as the inverse of the Peclet tensor of the flow at the
scale $n$ assuming that all the smaller scales have been completely averaged.

\begin{Definition}
We call "local renormalized Peclet tensor" $(\Pe^n)_{n\in \N}$ the inverse of the \emph{renormalization core}
\begin{equation}
\Pe^n=(A^n)^{-1}.
\end{equation}
\end{Definition}

\subsubsection{Pathologies of the \emph{renormalization core}.}

\begin{Definition}
We call stability of the \emph{renormalization core} \eref{wdkbkwjdebkljb81} the sequence
\begin{equation}
\lambda^-_n:=\inf_{0\leq p\leq n} \lambda_{\min}(A^p).
\end{equation}
We write
\begin{equation}
\lambda^-_\infty:=\lim_{n\rightarrow \infty} \lambda^-_n \quad \text{and}\quad \lambda^-:=\liminf_{n\rightarrow
\infty} \lambda_{\min}(A^n).
\end{equation}
The \emph{renormalization core} is said to be stable if and only if $\lambda^->0$.
\end{Definition}
\begin{Definition}\label{sdlikkdnkdnkd911}
We call anisotropic distortion of the \emph{renormalization core} \eref{wdkbkwjdebkljb81} the sequence
\begin{equation}
\mu_n=\sup_{0\leq p \leq n} \big(\lambda_{\max}(A^p)/\lambda_{\min}(A^p)\big).
\end{equation}
We write
\begin{equation}
\mu_\infty:=\lim_{n\rightarrow \infty} \mu_n \quad \text{and}\quad \mu:=\limsup_{n\rightarrow \infty}
\big(\lambda_{\max}(A^n)/\lambda_{\min}(A^n)\big).
\end{equation}
The \emph{renormalization core} \eref{wdkbkwjdebkljb81} is said to have unbounded (bounded) anisotropic distortion  if and only if $\mu=\infty$ ($\mu<\infty$).
\end{Definition}
\begin{Definition}
We call ubiety of  the \emph{renormalization core} \eref{wdkbkwjdebkljb81} the sequence
\begin{equation}
\lambda^+_n:=\sup_{0\leq p\leq n} \lambda_{\max}(A^p).
\end{equation}
We write
\begin{equation}
\lambda^+_\infty:=\lim_{n\rightarrow \infty} \lambda^+_n \quad \text{and} \quad \lambda^+:=\limsup_{n\rightarrow
\infty} \lambda_{\max}(A^n).
\end{equation}
The \emph{renormalization core}  is said to be vanishing if and only if $\lambda^+=0$
\end{Definition}
\begin{Definition}
The \emph{renormalization core} \eref{wdkbkwjdebkljb81} is said to be bounded if and only if $\lambda^+<\infty$.
\end{Definition}

The \emph{renormalization core} is gifted with remarkable properties which will be analyzed in details in
subsection \ref{sdhdgddzdg73tt6163a}. Before proceeding to super-diffusion we will give a first theorem
stressing the role of the stability of the \emph{renormalization core}, that is to say  the fact that the local
renormalized Peclet tensor stays bounded away from infinity. Indeed, it follows from  theorem
\ref{asjhsskbjs801} that the averaging paradigm for our model is valid if the \emph{renormalization core} is
stable, and has bounded anisotropic distortion. We may naturally wonder whether the fact that the local
renormalized Peclet tensor stays bounded away from infinity is sufficient, the answer is positive as shown by
the following theorem which will be proven in subsection \ref{sdjjddgjsdgd662fgs52121}.

\begin{Theorem}\label{djshdddsdbhdaaaas44aab881}
 Writing  $C= C_d K_0^2 (1-1/\gamma_{\min})^{-1}$ we have
\begin{enumerate}
\item  If the \emph{renormalization core} is not bounded ($\lambda^+=\infty$) then it is not stable ($\lambda^-=0$)
\item If the \emph{renormalization core} is stable ($\lambda^->0$) then it is bounded and
\begin{equation}
\lambda^+\leq  \frac{C}{\lambda^{-}}.
\end{equation}
\item The \emph{renormalization core} has unbounded anisotropic distortion ($\mu=\infty$) if and only if it is not stable ($\lambda^-=0$)
\item If the \emph{renormalization core} is stable ($\lambda^->0$) then  it has bounded anisotropic distortion ($\mu<\infty$) and
\begin{equation}
\mu \leq \frac{C}{(\lambda^-)^2}.
\end{equation}
\end{enumerate}
\end{Theorem}
Combining theorem \ref{djshdddsdbhdaaaas44aab881} and \ref{asjhsskbjs801} we obtain that if the
\emph{renormalization core} is stable then the local turbulent eddy conductivity diverges towards infinity like
$\gamma_n$ independently of the geometry of the eddies (if it is not stable the behavior of the local turbulent
eddy conductivity depends on the geometry of the eddies ). More precisely we have the following theorem.
\begin{Theorem}\label{asjhssswskbjs801}
Under hypotheses \hypoa, \hypob, \hypoc and \hypocf, if the \emph{renormalization core} is stable then there
exists $C$ such that for $\rho_{\min}^\alpha>C \gamma_{\max}$ one has
\begin{equation}
\begin{split}
\limsup_{n\rightarrow \infty} \frac{\ln\Big( \lambda_{\max} \big(\sigma_{sym}(\kappa I_d,\Gamma^{0,n})\big)\Big)}{\ln \gamma_n}\leq 1+ \epsilon
\end{split}
\end{equation}
and
\begin{equation}
\begin{split}
\liminf_{n\rightarrow \infty} \frac{\ln\Big( \lambda_{\min} \big(\sigma_{sym}(\kappa I_d,\Gamma^{0,n})\big)\Big)}{\ln \gamma_n}\geq 1- \epsilon
\end{split}
\end{equation}
with $\epsilon:= 0.5 \big(C\gamma_{\max}/(\rho_{\min})\big)^\frac{1}{2}<0.5$ and $C:=K_\alpha
h(d,K_0/\lambda^-)$.  $h$ being a finite increasing positive function in each of its arguments.
\end{Theorem}
\begin{Remark}
For a real flow, call $\sigma(r)$ the local turbulent diffusivity of the flow at the scale $r$ and $v(r)$ the
magnitude of the vector velocity field at that scale then the key relation implying that the distortions created
at the scale $r$ are dissipated by the mixing power of the smaller scales is the relation
\begin{equation}\label{dgdfdfdqqssqqs55}
\sigma(r)\sim r v(r)
\end{equation}
 This relation is at the core of the
Kolmogorov (K41) analysis and the analysis of fully developed turbulence by Landau-Lifschitz \cite{Lali84}. The
result given in  Theorem \ref{asjhssswskbjs801} corresponds to the relation \eref{dgdfdfdqqssqqs55} obtained and
used from a heuristic point of view (dimension analysis) by physicists.
\end{Remark}

\subsection{Super-diffusion}\label{sddduduzdd99827821ze61}
\paragraph{Anomalous fast exit times}
We write $\tau(r)$ the exit time of the process $y_t$ \eref{shsshkjjb81} from the ball $B(0,r)$. We write $\E_x$ the expectation associated to the process $y_t$ started from the point $x$. We write $\Vol\big(B(0,r)\big)$ the Lebesgue measure of $B(0,r)$. We define $n(r)$ as the number of (smaller) scales which will be considered as averaged at the scale $r$.
\begin{equation}
n(r):=\sup\{p\in \N\,:\, R_p\leq r\}.
\end{equation}
Let  $m_r$ be the Lebesgue probability measure  on the ball $B(0,r)$ defined by
\begin{equation}
m_r(dx):=\frac{dx}{\int_{B(0,r)}dx} 1_{B(0,r)}.
\end{equation}
We will consider  the mean exit time for the process started with initial distribution $m_{r}$, i.e.
\begin{equation}\label{shyqgsfgsxfydadhddaajhdgd788771s2qaqaajqqay}
\begin{split}
\E_{m_r}\big[\tau(r)\big]=\frac{1}{\Vol\big(B(0,r)\big)}\int_{B(0,r)}\E_x\big[\tau(r)\big]\,dx.
\end{split}
\end{equation}

\begin{figure}[htbp]
\centerline{\includegraphics[width=0.3\textwidth,height= 0.3\textwidth]{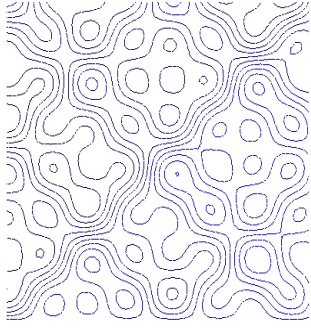}} \caption{Superposition of
cellular flows. \label{pavvvvcerticullll}}
\end{figure}

We will prove in subsection \ref{eefifgirfgzrg898731we} the following theorem.
\begin{Theorem}\label{hasgddfdgfwdysdxx7877761221y}
Under hypotheses \hypoa, \hypob, and \hypoc  with $\alpha=1$, if the \emph{renormalization core} is stable
($\lambda^->0$) then there exists a constant $Q$ such that for $\rho_{\min}>Q \gamma_{\max}$ one has
\begin{equation}\label{shyqgsfgsxfydadhddaajhdgd788771s2qaqaaqqay}
\begin{split}
\limsup_{r\rightarrow \infty}\frac{1}{\ln r} \ln\Big(\E_{m_r}\big[\tau(r)\big]\Big)<2.
\end{split}
\end{equation}
More precisely for $r>R_1$ one has
\begin{equation}\label{shyqgsfgsxfydadhddaajhdgd7887712qaqaaqqay}
\begin{split}
\E_{m_r}\big[\tau(r)\big]=r^{2-\nu(r)}
\end{split}
\end{equation}
with
\begin{equation}\label{skdhdgddzwzyuzdgz712}
\nu(r)=\frac{\ln \gamma_{n(r)}}{\ln r}\big(1+\epsilon(r)\big)+\frac{C(r)}{\ln r}.
\end{equation}
and  $|C(r)|\leq C(d,K_0,\gamma_{\max})+|\ln \lambda^-_\infty|$,
\begin{equation}
|\epsilon(r)| < 0.5 \big(\frac{Q \gamma_{\max}}{\rho_{\min}}\big)^{\frac{1}{2}} \leq 0.5
\end{equation}
and there exists  a finite increasing positive function in each of its arguments $F$ such that
\begin{equation}\label{dskdhdhggdjhgd873223} Q:=\frac{1}{(\ln \gamma_{\min})^2}F\big(d,\frac{(1+K_0)^2
(1-1/\gamma_{\min})^{-1}+\kappa}{\lambda^-_{\infty}}\big) (1+K_1).
\end{equation}
\end{Theorem}
\begin{Remark}
Equation \eref{skdhdgddzwzyuzdgz712} shows that the anomalous constant is directly related to the number of
effective scales. Observe that
\begin{equation}\label{skdhdxxaagddzwzavyuazdgz712}
\nu(r)\leq \frac{\ln \gamma_{\max}}{\ln \rho_{\min}}\Big(1+0.5 \big(\frac{Q
\gamma_{\max}}{\rho_{\min}}\big)^{\frac{1}{2}}\Big)+\frac{C(d,K_0,\gamma_{\max})}{\ln r}
\end{equation}
and
\begin{equation}\label{skaqdhdxxaagddzwzavyuzdgz712}
\nu(r)\geq \frac{\ln \gamma_{\min}}{\ln \rho_{\max}}\Big(1-0.5 \big(\frac{Q
\gamma_{\max}}{\rho_{\min}}\big)^{\frac{1}{2}}\Big)-\frac{C(d,K_0,\gamma_{\max})}{\ln r}
\end{equation}
and $\nu(r)>0.4 (\ln \gamma_{\min}/\ln \rho_{\max})$ for $r$ large enough. The anomalous parameter $\nu(r)$ is
not a constant because the model is not self-similar, in a self similar case
($\gamma_{\min}=\gamma_{\max}=\gamma$ and $\rho_{\min}=\rho_{\max}=\rho$) one would have at a logarithmic
approximation $$\E[\tau(r)]\sim r^{2-\nu} \quad \text{with} \quad \nu\sim \frac{\ln \gamma}{\ln \rho}.$$ The
error terms in $\nu(r)$ are explained by the interaction between the scales which are sensitive to the
particular geometry of the eddies. We remind that we consider a quenched model and it is not a priori obvious
that the transport should depend only on the power law in velocity field and not on its particular geometry.
\end{Remark}
Sufficient (and necessary) conditions for the stability of the \emph{renormalization core} ($\lambda^->0$) will be
given in subsection \ref{sdhdgddzdg73tt6163a}; we refer to theorems \ref{djshqadddsdbhdaaaabssd881},
\ref{djsdsdddswahdddsdbhdaaaab881} and \ref{sajssgszsgzs8712}. In particular
 if $d=2$ and if for all $k$, $E^k=E$ where $E$ corresponds to the cellular flow
 ($E_{12}(x,y):=\sin(2\pi x)\cos(2\pi y)$) then the renormalization core is stable ($\lambda^->0$).
 We have illustrated the contour lines of the superposition of $4$ scales of cellular flows in figure
  \ref{pavvvvcerticullll}.

 There exists an important literature on the fast transport phenomenon in turbulence addressed
 (from both heuristic and rigorous point of view) by using the tools of homogenization or renormalization; we refer
 to  \cite{MR82a:60149}, \cite{AvMa90}, \cite{AvMa94}, \cite{FuGl91}, \cite{Gli92},  \cite{GlZh92},
 \cite{Zh92}, \cite{GaKu95}, \cite{IsKa91},  \cite{Gau98}, \cite{Ave96},  \cite{Bha99}, \cite{FaKo01},
 \cite{BeOw00c}, \cite{MR2002c:60047}, \cite{AmFa02} and this panorama is far from being complete, we refer to
 \cite{MajKra99} and \cite{Woy00} for a survey.\\
For non exactly solvable models (non shear flows) asymptotic fast scaling in the transport behavior have been
obtained in the framework of spectral averaging in turbulence. Along this axis L. Piterbarg has obtained
\cite{MR98i:76046} fast asymptotic  scaling
 after averaging the transport with respect to the law of the velocity field and the thermal noise and
  rescaling with respect to space and time. More recently
S. Olla and T. Komorowski \cite{KoOl02} have observed the asymptotic anomalously fast
behavior of the mean squared displacement averaged with respect to the thermal noise, the law of the velocity field and time.
 A. Fannjiang \cite{Fa02} has studied a model where the law of separation of two particles is postulated to be the transport
  law of a single one as studied in \cite{KoOl02} and
  \cite{MR98i:76046}.

\paragraph{Fast mixing.}
In order to show that the phenomenon presented in theorem \ref{hasgddfdgfwdysdxx7877761221y} is super-diffusion
and not mere convection, we must compute the rate at which particles do separate and show that this rate follows
the same fast behavior. More precisely we will consider $(y_t,z_t)\in \R^{d}\times \R^d$ where $y_t$ is the
solution of \eref{shsshkjjb81} and $z_t$ follows the following stochastic differential equation
\begin{equation}\label{shsshkjjbaqa81ya}
dz_t=\sqrt{2 \kappa} d \bar{\omega}_t+\nabla .\Gamma(z_t)\,dt.
\end{equation}
Where $\bar{\omega}_t$ is a standard Brownian motion independent of $\omega_t$. Thus $y_t$ and $z_t$ can be seen as two particles transported by the same drift but with independent identically distributed  noise.
Let us write $B(0,r,l)$ the following subset of $\R^d \times \R^d$
\begin{equation}
B(0,r,l):=\big\{(y,z)\in \R^d\times \R^d\,:\, |y-z|<r \quad \text{and}\quad y^2+z^2< l^2 \big\}.
\end{equation}
We write $\E_{y,z}\big[\tau(r,l)\big]$ the expectation of the exit time of the diffusion $(y_t,z_t)$ from $B(0,r,l)$ with $(y_0,z_0)=(y,z)$. Let  $m_{r,l}$ be the Lebesgue probability measure  on the set $B(0,r,l)$ defined by
\begin{equation}
m_{r,l}(dy\,dz):=\frac{dy\,dz}{\int_{(y,z)\in B(0,r,l)}dy\,dz}1_{B(0,r,l)}.
\end{equation}
We will consider  the mean exit time for the process $(y_t,z_t)$ started with initial distribution $m_{r,l}$, i.e.
\begin{equation}\label{shyqgsfgsxfaqydadhddaajhdgd788771s2qaqaaqqay}
\begin{split}
\E_{m_{r,l}}\big[\tau(r,l)\big]=\frac{1}{\Vol\big(B(0,r,l)\big)}\int_{B(0,r,l)}\E_{y,z}\big[\tau(r,l)\big]\,dx.
\end{split}
\end{equation}
We have the following theorem proven in subsection \ref{eefifgirfgzrg898731we}.
\begin{Theorem}\label{hasgddfdgfwasdsd7q877761221by}
Under hypotheses \hypoa, \hypob, and \hypoc  with $\alpha=1$, if the \emph{renormalization core} is stable then
there exists a constant $Q$ such that for $\rho_{\min}>Q \gamma_{\max}$ one has
\begin{equation}\label{shyqgsfgsxfydadhddaajhdgd78j8771s2qaqaaqqay}
\begin{split}
\limsup_{r\rightarrow \infty}\lim_{l\rightarrow \infty} \frac{1}{\ln r} \ln \Big( \E_{m_{r,l}}\big[\tau(r,l)\big]\Big)<2.
\end{split}
\end{equation}
More precisely for $r>R_1$ one has
\begin{equation}\label{shyqgsfgsxfydaadhddaajhadgd7887712qaqaaqqay}
\begin{split}
\lim_{l\rightarrow \infty} \E_{m_{r,l}}\big[\tau(r,l)\big]=r^{2-\nu(r)}.
\end{split}
\end{equation}
where $\nu(r)$ is given by \eref{skdhdgddzwzyuzdgz712} and $Q$ by \eref{dskdhdhggdjhgd873223}.
\end{Theorem}
\begin{Remark}
It is easy to extend this theorem to any finite number of particles driven by the same flow but independent
thermal noise.
\end{Remark}

\paragraph{Strong self-averaging property.}
A trivial consequence of theorem \ref{hasgddfdgfwdysdxx7877761221y} and \ref{hasgddfdgfwasdsd7q877761221by} is the fact that fast mixing is an almost sure event. More precisely,
let us write $H(r)$ and $H(r,l)$ the events
$$
H(r):=\Big\{\tau(r) \leq r^{2-\delta}  \Big\}\quad \text{and}\quad H(r,l):=\Big\{\tau(r,l) \leq r^{2-\delta}
\Big\}
$$
with $\delta=0.9 \ln \gamma_{\min}/\ln \rho_{\max}$. Observe that $\delta>0$ and we have the following theorem
\begin{Theorem}\label{sjdhdjgdd87237sg731}
Under hypotheses \hypoa, \hypob and \hypoc  with $\alpha=1$, if the \emph{renormalization core} is stable then
there exists a constant $Q$ such that for $\rho_{\min}> Q \gamma_{\max}$ one has
\begin{equation}
\lim_{r\rightarrow \infty} \P_{m_{r}}\big[H(r)\big]=1\quad \text{and}\quad \lim_{r\rightarrow \infty}
\lim_{l\rightarrow \infty}\P_{m_{r,l}}\big[H(r,l)\big]=1.
\end{equation}
\end{Theorem}
In this theorem $Q$ is given by \eref{dskdhdhggdjhgd873223}, by slightly modifying the constants.

\subsection{Diagnosis of  \emph{renormalization core}'s pathologies}\label{sdhdgddzdg73tt6163a}
With subsection \ref{sddduduzdd99827821ze61} we have seen that our model is super-diffusive if the \emph{renormalization core} is stable.
With this  subsection  we will give necessary and sufficient conditions for the stability of the \emph{renormalization core} by
 analyzing in details its dynamic. The results given here will be proven in subsection \ref{sdjjddgjsdgd662fgs52121}.

\paragraph{Diffusive properties of the eddies at vanishing molecular conductivity.}
We will need the following functions $V$ and $W$ describing the effective behavior of the eddies $E^k$ of the
\emph{renormalization core} at vanishing molecular conductivity. For $E\in \SL(\T^d)$ and $\zeta>0$ we write
\begin{equation}
V(\zeta,E):=\frac{\lambda_{\min}\big(\sigma_{\sym}(\zeta I_d,E)\big)}{\zeta}.
\end{equation}
Observe that by the variational formulation \eref{HoCoeDNorPre12} one has
\begin{equation}\label{gggrrrr33221}
1/V(\zeta,E)=\sup_{\xi\in\Sp^{d-1}}  \inf_{(f,H)\in C^\infty(\T^d)\times \S(\T^d)}\int_{\T^d}|\xi-\nabla .H+
E\nabla f|^2\,dx + \zeta^2 \int_{\T^d}|\nabla f(x)|^2\,dx.
\end{equation}
Where we have written $\Sp^{d-1}$ the unit sphere of $\R^d$ centered on $0$.
Observe that if for $\xi \in \Sp^{d-1}$, $\nabla .E.\xi \not\equiv 0$ one has $\forall \zeta >0$ $V(\zeta,E)>1$. Moreover $V$ is continuous and decreasing in $\zeta$.
Let us define
\begin{Definition}
\begin{equation}
V(\zeta):= \inf_{n\in \N} V(\zeta,E^n)
\end{equation}
\begin{equation}\label{sjsbjbdjbd9811}
V(0):=\lim_{\zeta \downarrow 0} V(\zeta)
\end{equation}
\end{Definition}
Observe that $V(\zeta)$ is a decreasing function in $\zeta$ thus the limit \eref{sjsbjbdjbd9811} is well defined and belongs to $[1,\infty]$. We define for $x\in \big(1,V(0)\big)$, the inverse function $V^{-1}$ as
\begin{equation}
V^{-1}(x):=\sup \{y >0\,:\, V(y)>x\}.
\end{equation}
Observe that if $V(0)>1$, $V^{-1}(x)$ is a decreasing function of $x$ in $\big(1,V(0)\big)$.\\
Similarly we introduce
\begin{equation}
W(\zeta,E):=\frac{\lambda_{\max}\big(\sigma_{\sym}(\zeta I_d,E)\big)}{\zeta}.
\end{equation}
Observe that by the variational formulation \eref{HoCoeDNorPre14} one has
\begin{equation}\label{sshbskjhbcbb98811}
W(\zeta, E)=1+\sup_{l\in \S^{d-1}}\inf_{\xi \perp l, (f,H) \in C^\infty(\T^d)\times
\S(\T^d)}\zeta^{-2}\int_{\T^d}|\xi-\nabla .H-E(l-\nabla f)|^2\,dx+\int_{\T^d}|\nabla f|^2\,dx.
\end{equation}
Observe that if for $l \in \Sp^{d-1}$, $\nabla .E.l \not\equiv 0$ one has $\forall \zeta >0$ $W(\zeta,E)>1$. Moreover $W$ is continuous and decreasing in $\zeta$. Let us define
\begin{Definition}
\begin{equation}
W(\zeta):= \sup_{n\in \N} W(\zeta,E^n)
\end{equation}
\begin{equation}\label{sjsbjbdjddcefcbd9811}
W(0):=\lim_{\zeta \downarrow 0} W(\zeta)
\end{equation}
\end{Definition}
Observe that $W(\zeta)$ is a decreasing function in $\zeta$ thus the limit \eref{sjsbjbdjddcefcbd9811} is well defined and belongs to $[1,\infty]$.\\
We remind that for $E\in \SL(\T^d)$ and $\zeta>0$, one has
\begin{equation}
1\leq V(\zeta,E)\leq 1+ C_{d}\zeta^{-2} \lambda_{\min}\Big(\int_{\T^d}{^tE(x)E(x)}\,dx \Big)
\end{equation}
and
\begin{equation}
1\leq W(\zeta,E)\leq 1+ C_{d}\zeta^{-2} \lambda_{\max}\Big(\int_{\T^d}{^tE(x)E(x)}\,dx \Big).
\end{equation}
Moreover the behavior of $V(\zeta,E)$ and $W(\zeta,E)$ at vanishing molecular conductivity (as $\zeta \downarrow
0$) and their connections with the stream lines of the eddies has been widely studied in the literature (we
refer to \cite{IsKa91}, \cite{FaPa94} and the references inside). Thus it has been obtained \cite{FaPa94} that
for any $\beta \in [-\frac{1}{2},0]$ there exist $E\in \SL(\T^d)$ such that $V(\zeta,E)=W(\zeta,E)$ and as
$\zeta \downarrow 0$
\begin{equation}
 V(\zeta,E))\sim c^* \zeta^\beta.
\end{equation}
Where $c^*$ can be calculated explicitly in several cases. An particular example with  $V(\zeta,E))\sim -c^* \ln \zeta$
is also given in \cite{FaPa94}. For anisotropic cases, for any $\delta \in [0,1/2)$ there exist $E\in \SL(\T^d)$ such that \cite{FaPa94}
\begin{equation}
W(\zeta,E)\sim c^*_1 \zeta^{3\delta -2} \quad \quad V(\zeta,E)\sim c^*_2 \zeta^{-\delta}.
\end{equation}

\paragraph{The stability of the \emph{renormalization core} and its anisotropy.}
 Theorem \ref{djshddsadsdbhdaaaab881} shows that the anisotropy of the local turbulent conductivity is one of the causes of the instability of the \emph{renormalization core}. It is natural to wonder whether the converse is true, the answer is positive at low flow rate as shown by the following theorem and corollary.
\begin{Theorem}\label{djshddsadsdbhdaaaab881}
If the \emph{renormalization core} has bounded anisotropic distortion ($\mu<\infty$) then
\begin{enumerate}
\item  if $\gamma_{\max}<V(0)$ then the \emph{renormalization core} is stable ($\lambda^->0$). Moreover if the monotony of $V$ is strict then
\begin{equation}\label{ddkdjbkdjbdjkdbdjd98987101}
\lambda^- \geq \mu^{-\frac{1}{2}} \frac{V^{-1}(\gamma_{\max})}{\gamma_{\max}}.
\end{equation}
\item If  $W(0)<\infty$ then the \emph{renormalization core} is bounded from above and
\begin{equation}\label{ddkdjbkdjbdjkdbdjd98987101q}
\lambda^+ \leq  \mu^\frac{1}{2} C_d K_0(\gamma_{\min}-1)^{-\frac{1}{2}} \big(1+W(0)\big).
\end{equation}
\end{enumerate}
\end{Theorem}
\begin{Corollary}\label{djshddddddsadsdbhdaaaab881}
If $\gamma_{\max}<V(0)$  and the renormalization core is not stable ($\lambda^-=0$) then it has  unbounded
anisotropic distortion ($\mu=\infty$).
\end{Corollary}

\begin{Definition}
The flow is said to be isotropic if for all $\zeta>0$, $k\in \N$, $\sigma_{\sym}(\zeta I_d,E^k)$ is a multiple of the identity matrix.
\end{Definition}

\begin{Definition}
The \emph{renormalization core} is said to be isotropic if for all $k$, $A^k$ is a multiple of the identity matrix. We then write $A^k=\lambda(A^k) I_d$.
\end{Definition}
Observe that if the flow is isotropic then so is the \emph{renormalization core}, $\mu=1$ and
 from theorem \ref{djshddsadsdbhdaaaab881} we obtain the following corollary.
\begin{Corollary}
If the flow is isotropic then the \emph{renormalization core} is stable for $\gamma_{\max}< V(0)$.
\end{Corollary}
Combining theorem \ref{djshddsadsdbhdaaaab881} and \ref{djshdddsdbhdaaaas44aab881} we obtain that for $\gamma_{\max}<V(0)$ the \emph{renormalization core} is not stable if and only if it has unbounded anisotropic distortion. Moreover we have
the following theorem
\begin{Theorem}\label{djshqadddsdbhdaaaabssd881}
 If the \emph{renormalization core} has bounded anisotropic distortion, the monotony of $V$ is strict, $\gamma_{\max}<V(0)$  then the \emph{renormalization core} is stable ($\lambda^->0$) and
\begin{equation}\label{ddkdjsswbkdqyjbdaqjkdbdjd98987101}
C_1 \leq \lambda^- \lambda^+ \leq C_2.
\end{equation}
with $C_1=\big(V^{-1}(\gamma_{\max})/\gamma_{\max}\big)^2$ and $C_2=C_d K_0^2 (1-1/\gamma_{\min})^{-1}$.
\end{Theorem}
We believe that equation \eref{ddkdjsswbkdqyjbdaqjkdbdjd98987101} could be at the origin of the isotropy of
turbulence at small scales.  Let us observe that if $V(0)=\infty$ then the stability of the
\emph{renormalization core} is equivalent to the fact that it has bounded anisotropic distortion. It is easy to
build from theorem \ref{djshqadddsdbhdaaaabssd881} and the analysis of $V(\zeta,E)$ given above,  examples of
flows with stable \emph{renormalization core} and thus a strongly-super-diffusive behavior. In particular  we
have the following theorem
\begin{Theorem}\label{djsdsdddswahdddsdbhdaaaab881}
If $V(0)=\infty$ and the \emph{renormalization core} is isotropic then  under hypotheses \hypoa, \hypob and
\hypoc
 with $\alpha=1$ the flow is strongly super-diffusive for $\rho_{\min}>11 Q \gamma_{\max}$ (where
$Q$ is given by \eref{dskdhdhggdjhgd873223}) and theorems \ref{hasgddfdgfwdysdxx7877761221y},
\ref{hasgddfdgfwasdsd7q877761221by} and \ref{sjdhdjgdd87237sg731} are valid (with $\lambda^->0$).
\end{Theorem}
Observe that if $d=2$ and if for all $k$, $E^k=E$ where $E$ corresponds to the cellular flow
($E_{12}(x,y):=\sin(2\pi x)\cos(2\pi y)$) then $V(0)=\infty$ (\cite{FaPa94}) and the renormalization core is
stable ($\lambda^->0$).

\paragraph{Viscosity implosion.}

It is easy to obtain that if there exists $\delta >0$ such that for all $k$ the drift $\nabla .E^k$ is null on
$[0,1]^d\setminus[\delta,1-\delta]^d$ then $W(0)<\infty$. Moreover we have the following theorem.
\begin{Theorem}\label{djsswahdddsdbhdaaaab881}
 If $\gamma_{\min}>W(0)$ then the \emph{renormalization core} is vanishing  with exponential rate and
\begin{equation}
\limsup_{n\rightarrow\infty}\frac{\ln\big(\lambda_{\max}(A^n)\big)}{n}\leq \ln\big(\frac{W(0)}{\gamma_{\min}}\big).
\end{equation}
\end{Theorem}
It follows from theorem \ref{djsswahdddsdbhdaaaab881} the \emph{renormalization core} can be isotropic and not stable at the same time. Now it is natural to wonder whether a  \emph{renormalization core} (and thus the transport properties of the flow) may undergo a brutal  alteration.
\begin{Definition}
We call \emph{viscosity implosion} the bifurcation from a stable \emph{renormalization core} to a vanishing \emph{renormalization core}.
\end{Definition}
We will now analyze this phenomenon.
\begin{Definition}
The flow is said to be self-similar if and only if $\gamma_{\max}=\gamma_{\min}=\gamma$,
$\rho_{\max}=\rho_{\min}=\rho$ and for all $k$, $E^k=E^0=E$.
\end{Definition}
Let us remind  that a real turbulent flow has a non self-similar multi-scale structure, we refer to
\cite{DiCa97}. Observe that if the flow is self-similar then $V(\zeta)=W(\zeta)$. In this case we will write
\begin{equation}
\gamma_c:=V(0)
\end{equation}

\begin{Theorem}\label{sajssgszsgzs8712}
Assume the flow to be self-similar and isotropic
\begin{enumerate}
\item If $\gamma <\gamma_c$ then the \emph{renormalization core} is stable ($\lambda^->0$) and
\begin{equation}\label{dskdsudsuhdusd}
\lim_{n\rightarrow \infty} A^n=\zeta_0 I_d.
\end{equation}
where $\zeta_0$ is the unique solution of $V(\zeta_0)=\gamma$ \item If $\gamma =\gamma_c$ and
$(V(0)-V(x))x^{-p}$ admits a non null limit as $x\downarrow 0$ with $p>0$ then the \emph{renormalization core}
is vanishing  with polynomial rate (in particular $\lambda^+=\lambda^-=0$):
\begin{equation}
\lim_{n\rightarrow \infty}  \frac{\ln \lambda(A^n)}{\ln n} =-\frac{1}{p}.
\end{equation}
\item If $\gamma > \gamma_c$ then the \emph{renormalization core} is vanishing with exponential rate (in
particular $\lambda^+=\lambda^-=0$)
\begin{equation}
\lim_{n\rightarrow \infty} \frac{1}{n}\ln \lambda (A^n)=\ln\big(\frac{\gamma_c}{\gamma} \big).
\end{equation}
\end{enumerate}
\end{Theorem}
\begin{figure}[htbp]
  \begin{center}
   \subfigure[An implosive
eddy geometry.\label{partdddccccddicullll}]
               {\includegraphics[width=0.3\textwidth,height= 0.3\textwidth]{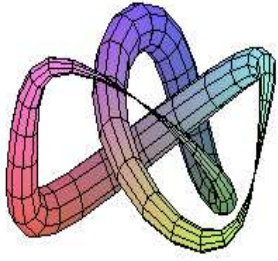}}
    \goodgap
    \subfigure[Stable \emph{renormalization core}.\label{figfffrrr}]
          {\includegraphics[width=0.3\textwidth,height= 0.3\textwidth]{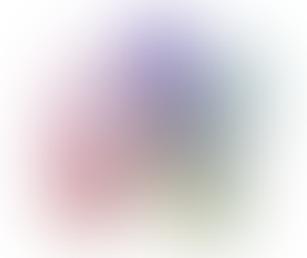}}
    \goodgap
   \subfigure[Vanishing \emph{renormalization core}.\label{figdjdueAfirstB}]
        {\includegraphics[width=0.3\textwidth,height= 0.3\textwidth]{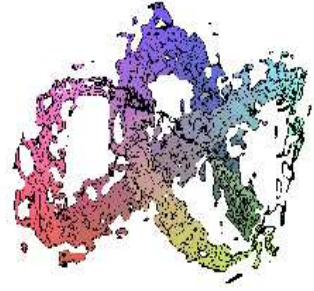}}\\[-10pt]
    \caption{Viscosity implosion.}
    \label{figyyassAfirst}
\end{center}
\end{figure}
It follows from equation \eref{dskdsudsuhdusd} that if the flow is self-similar and the renormalization core isotropic and $E$ non constant then $V(0)>1$ and for $1<\gamma<V(0)$ the flow is strongly super-diffusive and theorems \ref{hasgddfdgfwdysdxx7877761221y}, \ref{hasgddfdgfwasdsd7q877761221by} and \ref{sjdhdjgdd87237sg731} are valid (with $\lambda^->0$).\\
The viscosity implosion of the \emph{renormalization core} implies that the strong self-averaging property of
the flow collapse towards a chaotic coupling between the scales. Let us give a particular example to illustrate
what we mean by such bifurcation. The flow is assumed to be self-similar and isotropic and the stream lines of
the eddy $E$ over a period $[0,1]^3$ are given in the figure \ref{partdddccccddicullll}. Since there exists
$\delta >0$ such that  the drift $\nabla .E$ is null on $[0,1]^d \setminus [\delta,1-\delta]^d$ we have
$\gamma_c<\infty$ with the eddy illustrated in figure \eref{partdddccccddicullll}. Now imagine that one puts a
drop of dye in such a flow and observe its transport at very large spatial scale. We have illustrated in the
figure \ref{figyyassAfirst} a metaphorical illustration of what one could see, it would be interesting to run
numerical simulations to analyze the behavior of a drop of dye at the transition between a stable and vanishing
renormalization core. For $\gamma<\gamma_c$ dye is transported by strong super-diffusion, and the density of its
colorant in the flow is homogeneous (figure \ref{figfffrrr}). Moreover in the domain $(0,\gamma_c)$ an increase
of the flow rate $\gamma$ in the eddies is compensated by an increase of the diffusive (dissipative) power of
the smaller eddies. The picture undergoes a brutal transformation at $\gamma\geq \gamma_c$; in this domain an
increase of $\gamma$ results in the growth of the advective power of the eddies but their diffusive power
remains bounded and can no longer compensate convection. The diffusive power of the smaller scales becomes
dominated by the convective power of the eddy at the observation scale (figure \ref{figdjdueAfirstB}). The drop
dye is then transported by advection and presents high density gradients.

\paragraph{Variational formulae for $\gamma_c$.}
Assume the flow to be self-similar and isotropic. Thus from equation \eref{gggrrrr33221} it is easy to obtain
that
\begin{equation}\label{gshsqqwwqgfs111}
\gamma_c^{-1}= \inf_{(f,H)\in C^\infty(\T^d)\times \S(\T^d)}\int_{\T^d}|\xi-\diiv H+ E\nabla f|^2\,dx
\end{equation}
from equation \eref{sshbskjhbcbb98811} it is also easy to obtain that for any unit vector $l$ in $\R^d$
\begin{equation}\label{gshsgfs1qwaz12}
\gamma_c=1+\lim_{\zeta \downarrow 0} \inf_{\xi \perp l, (f,H) \in C^\infty(\T^d)\times
\S(\T^d)}\zeta^{-2}\int_{\T^d}|\xi-\diiv H-E(l-\nabla f)|^2\,dx+\int_{\T^d}|\nabla f|^2\,dx.
\end{equation}
Write $G$ the set of $f\in H^1(\T^d)$ such that there exists $\xi, l \in \R^d$ and $H$ a skew symmetric matrix
with coefficients in $H^1(\T^d)$ with $\xi \perp l$ and
\begin{equation}\label{gshsgfs1qwssqzwaz12}
\xi-\diiv H-E(l-\nabla f)=0. \end{equation}
 Then if $G=\emptyset$ it is easy to obtain from \eref{gshsgfs1qwaz12} that
$\gamma_c=\infty$. If $G\not= \emptyset$  then one has
\begin{equation}\label{gshsgfwwvs1qwaz12}
\gamma_c=1+ \inf_{f \in G}\int_{\T^d}|\nabla f|^2\,dx.
\end{equation}
The equation \eref{gshsgfs1qwssqzwaz12} is degenerate, thus it is not easy to prove a solution for that equation
in a general case and actually most of the time it has no solution which means  that $\gamma_c=\infty$. It would
be interesting to obtain non trivial criteria ensuring the existence of a solution for
\eref{gshsgfs1qwssqzwaz12}. The most trivial example of a stream matrix $E$ such that $\gamma_c<\infty$ is the
following one. Take $d=2$ and $E$ a skew symmetric matrix with $E_{1,2}=h$ where $h$ over the period $[0,1]^2$
is equal to
\begin{equation}
h(x_1,x_2)=\sin(2\pi x_1)\sin(2\pi x_2) g\Big(4 \big((x_1-0.5)^2 + (x_2-0.5)^2\big)\Big)
\end{equation}
where $g$ is any smooth function on [0,1] such that $g=1$ on $[0,1/3]$ and $g=0$ on $[2/3,1]$. Then it is easy
to check that $1<\gamma_c<\infty$  and estimate it from the variational formulae given above. For instance write
$G'$ the set of smooth $\T^d$ periodic function $f$  such that $\nabla f = e_2$ on $\{x\,:\,(x_1-0.5)^2 +
(x_2-0.5)^2\leq 1/6 \}$ then it is easy to check that
\begin{equation}\label{gshsgfs11ss2}
\gamma_c\leq 1+ \inf_{f\in G'}\int_{\T^d}|\nabla f|^2\,dx.
\end{equation}

\paragraph{The \emph{renormalization core} with a finite number of scales.}
The results given above were related to the asymptotic behavior of the \emph{renormalization core}. When the flow has only a finite number of scales we will give below quantitative estimates controlling the \emph{renormalization core}.
\begin{Theorem}\label{djshddaqadsdbhdb881}
The ubiety of the \emph{renormalization core} is bounded from above by the inverse of its stability.
Writing  $C= C_d K_0^2 (1-1/\gamma_{\min})^{-1}$ we have
$$
\lambda^+_n\leq \kappa+ \frac{C}{\lambda^{-}_{n-1}} \quad \text{and}\quad \mu_n\leq
\frac{\kappa}{\lambda^{-}_{n}}+ \frac{C}{(\lambda^{-}_{n})^2}
$$
\end{Theorem}

\begin{Theorem}\label{sudjzdd78727hg712}
We have
\begin{equation}\label{ddkdjbkdjbdjkdbdjad98987101}
\lambda^-_n \geq \min\Big(\lambda_{\min}(\kappa),(\mu_n)^{-\frac{1}{2}}
\frac{V^{-1}(\gamma_{\max})}{\gamma_{\max}} \Big).
\end{equation}
\end{Theorem}

\begin{Theorem}\label{dkdccizueztU721yys}
We have for $n\in \N$
\begin{equation}\label{ddkdjbkdjbdjkdbsdjd98987101q}
\lambda^+_n \leq \max\Big(\lambda_{\max}(\kappa),C_d K_0 \mu^\frac{1}{2}_n (\gamma_{\min}-1)^{-\frac{1}{2}} \Big) \big(1+W(0)\big).
\end{equation}
\end{Theorem}

In particular, observe that if $\gamma_{\max}<\infty$ and $\lambda^-=0$ then the stability of the
\emph{renormalization core} should decrease according to the following relation
 $\ln \lambda^-_n \sim -0.5 \ln \mu_n$ and its ubiety should increase like $\ln \lambda^+_n \sim 0.5 \ln \mu_n$.

\section{Proofs}\label{sdkjjjkhdh98ehh8e331}
\subsection{Averaging with two scales: proof of theorem \ref{0kjhhbxdhbo1222}}\label{skndkndkdndn8981111}
There are two strategies to prove theorem \ref{0kjhhbxdhbo1222}; the first one is based on the   relative
translation method introduced in \cite{Owh01}  and the  variational formulations of the effective conductivity;  this is the strategy used in \cite{BeOw00b}. The second one is new and based only on the relative translation method. Although the first strategy in the case considered here would give (the proof is rather long) a sharper estimate of the error term: $(1+\epsilon(R))^2$ with $\epsilon(R)= \frac{\|K\|_\alpha}{R^\alpha \lambda_{\min}(a) } f(d,\frac{\|a+P\|_\infty}{\lambda_{\min}(a)})$ instead of \eref{dkjdjbj8981111} we
 have preferred to write here the second one for its simplicity and the fact that it allows to obtain a lower and an upper bound at once without the need of any variational formulation. Let us now give this new alternative strategy.\\
By the variational formulation \eref{HoCoeDNorPre14} the effective conductivity $\sigma_{\sym}(a,S_R P+K)$ is continuous in $L^\infty(\T^d)$ norm with respect to the stream matrices $P$ and $K$ and by density it is sufficient to prove the estimate \eref{dkjdjbj8981111}
 assuming that $P$ and $K$ are smooth and belongs to $\S(\T^d)$.\\
First we will prove the following proposition where we have used the notation introduced in section \ref{djgddffdgdd7766121} (we write $E:=S_R P+K$).
\begin{Proposition}\label{dkdhdhbdhdbh89811h88122}
Let  $l\in \R^d$, $l\not=0$.
\begin{equation}\label{sdjdhdhhdwdcdfre223}
\bigg[ \Big({^tl}\sigma_{\sym}(a, S_R P+K)l\Big)^{\frac{1}{2}}-\Big({^tl}\sigma_{\sym}\big(\sigma_{\sym}(a,P),K\big)l\Big)^{\frac{1}{2}}\bigg]^2\leq J_1+J_2+J_3+J_4
\end{equation}
with
\begin{equation}\label{dkjdbjbffbuu98981122}
\begin{split}
J_1=\int_{(x,y)\in\T^d \times [-\frac{1}{2},\frac{1}{2}]^d} \nabla F^{a,E}_l(x+\frac{y}{R})\big(a-E(x+\frac{y}{R})\big) \nabla \chi^{a,P}(Rx+y)\nabla F^{\sigma_{\sym}(a,P),K}_l(x)\,dx\,dy
\end{split}
\end{equation}
\begin{equation}\label{sajsskjbsj9sj991}
\begin{split}
J_2=-\int_{\T^d \times [-\frac{1}{2},\frac{1}{2}]^d} \nabla F^{a,E}_l(x+\frac{y}{R})\big((a+P(Rx+y))\nabla F^{a,P}(Rx+y)-\sigma_{\sym}(a,P)\big) \nabla F^{\sigma_{\sym}(a,P),K}_l(x)\,dx\,dy
\end{split}
\end{equation}
\begin{equation}\label{dljsdbjbkjsbdkb8991}
\begin{split}
J_3=-\int_{\T^d \times [-\frac{1}{2},\frac{1}{2}]^d}\nabla F^{a,E}_l(x+\frac{y}{R})\big(K(x+\frac{y}{R})-K(x)\big)\nabla F^{a,P}(Rx+y)\nabla F^{\sigma_{\sym}(a,P),K}_l(x)\,dx\,dy
\end{split}
\end{equation}
\begin{equation}\label{skhsvhsjshh787711}
\begin{split}
J_4=\int_{\T^d \times [-\frac{1}{2},\frac{1}{2}]^d}\nabla F^{a,E}_l(x+\frac{y}{R})K(x)\nabla \chi^{a,P}(Rx+y)\nabla F^{\sigma_{\sym}(a,P),K}_l(x)\,dx\,dy
\end{split}
\end{equation}
\end{Proposition}
\begin{proof}
 Let us write
\begin{equation}
I=\int_{(x,y)\in\T^d \times [-\frac{1}{2},\frac{1}{2}]^d} \nabla F^{a,E}_l(x+\frac{y}{R})a \nabla F^{a,P}(Rx+y)\nabla F^{\sigma_{\sym}(a,P),K}_l(x)\,dx\,dy.
\end{equation}
Using Cauchy-Schwartz inequality and the formula \eref{dskjsdbjdb89018} one obtains that
\begin{equation}\label{sdjdhdhhdwdcdfre221}
I\leq \Big({^tl}\sigma_{\sym}(a, S_R P+K)l\Big)^{\frac{1}{2}}\Big({^tl}\sigma_{\sym}\big(\sigma_{\sym}(a,P),K\big)l\Big)^{\frac{1}{2}}.
\end{equation}
Now, writing $E=S_R P + K$ observe that
\begin{equation}\label{sdjdhdhhdwdcdfre222}
I=(I_1+I_2)/2
\end{equation}
 with
\[
\begin{split}
I_1=\int_{(x,y)\in\T^d \times [-\frac{1}{2},\frac{1}{2}]^d} \nabla F^{a,E}_l(x+\frac{y}{R})\big(a+E(x+\frac{y}{R})\big) \nabla F^{a,P}(Rx+y)\nabla F^{\sigma_{\sym}(a,P),K}_l(x)\,dx\,dy
\end{split}
\]
and
\[
\begin{split}
I_2=\int_{(x,y)\in\T^d \times [-\frac{1}{2},\frac{1}{2}]^d} \nabla F^{a,E}_l(x+\frac{y}{R})\big(a-E(x+\frac{y}{R})\big) \nabla F^{a,P}(Rx+y)\nabla F^{\sigma_{\sym}(a,P),K}_l(x)\,dx\,dy.
\end{split}
\]
Using
\[
\begin{split}
\int_{(x,y)\in\T^d \times [-\frac{1}{2},\frac{1}{2}]^d} \nabla F^{a,E}_l(x+\frac{y}{R})\big(a-E(x+\frac{y}{R})\big)l= {^tl}\sigma_{\sym}(a, S_R P+K)l
\end{split}
\]
and the fact that $\nabla F^{a,E}_l(x+\frac{y}{R})\big(a-E(x+\frac{y}{R})\big)$ is a divergence free vector field one obtains that
\[
\begin{split}
I_2=&{^tl}\sigma_{\sym}(a, S_R P+K)l-J_1
\end{split}
\]
with $J_1$ given by \eref{dkjdbjbffbuu98981122}.
Moreover
\begin{equation}
\begin{split}
I_1=G_0-J_2-J_3-J_4
\end{split}
\end{equation}
with
\begin{equation}
\begin{split}
G_0=&\int_{(x,y)\in\T^d \times [-\frac{1}{2},\frac{1}{2}]^d} \nabla F^{a,E}_l(x+\frac{y}{R})\big(\sigma_{\sym}(a,P)+K(x)\big)\nabla F^{\sigma_{\sym}(a,P),K}_l(x)\,dx\,dy
\\=&{^tl}\sigma_{\sym}\big(\sigma_{\sym}(a,P),K\big)l
\end{split}
\end{equation}
where we have used in the last equality the fact that $\big(\sigma_{\sym}(a,P)+K(x)\big)\nabla F^{\sigma_{\sym}(a,P),K}_l(x)$ is divergence free. And $J_2$, $J_3$, $J_4$ are given by \eref{sajsskjbsj9sj991}, \eref{dljsdbjbkjsbdkb8991} and \eref{skhsvhsjshh787711}.
Thus combining \eref{sdjdhdhhdwdcdfre221} and \eref{sdjdhdhhdwdcdfre222} we have obtained \eref{sdjdhdhhdwdcdfre223},
 which proves the proposition.
\end{proof}
Now we will show that $J_1$, $J_2$, $J_3$ and $J_4$ act as  error terms in the homogenization process.\\
Using $\operatorname{div}\big( (a+E)\nabla F^{a,E}_l\big)=0$  and observing that, $\nabla \chi^{a,P}(Rx+y)=\nabla_y \chi^{a,P}(Rx+y) $ and integrating by parts in $y$ one obtains  (writing $\partial^i([-\frac{1}{2},\frac{1}{2}]^d)=\{x\in [-\frac{1}{2},\frac{1}{2}]^d\,:\,x_i=-\frac{1}{2}\}$)
\begin{equation}\label{dslnlddndklndk990091}
J_1=G_1+G_2
\end{equation}
 with (writing $(e_i)_{1\leq i \leq d}$ the orthonormal basis of $\R^d$ compatible with the axis of periodicity of $\T^d$)
\begin{equation}\label{HoCoeDNo1zfgscv2864}
\begin{split}
G_1=&\sum_{i=1}^d\int_{(x,y^i)\in \T^d\times \partial^i([-\frac{1}{2},\frac{1}{2}]^d)}{^t\big(\nabla F^{a,E}_l(x+(y^i+e_i)/R)-\nabla F^{a,E}_l(x+y^i/R)\big)}(a-P(Rx+y^i)).e_i \\&\chi^{a,P}(Rx+y^i)\nabla F^{\sigma_{\sym}(a,P),K}_l(x) \,dx\,dy^i
 \end{split}
\end{equation}
\begin{equation}\label{HoCoeDNo1zfgscv2865}
\begin{split}
G_2=-&\sum_{i=1}^d\int_{(x,y^i)\in \T^d\times \partial^i([-\frac{1}{2},\frac{1}{2}]^d)}\big({^t\nabla F^{a,E}_l(x+(y^i+e_i)/R)}K(x+(y^i+e_i)/R)\\&-{^t\nabla F^{a,E}_l(x+y^i/R)}K(x+y^i/R)\big).e_i \chi^{a,P}(Rx+y^i)\nabla F^{\sigma_{\sym}(a,P),K}_l(x) \,dx\,dy^i.
 \end{split}
\end{equation}
Now we will need the following lemma which says that the solution of the two-scale cell problem keeps in its structure a signature of the fast period.
\begin{Lemma}\label{dslkdnlkcvnnf91112}
For $i\in \{1,\ldots d\}$ one has
\begin{equation}\label{HoCoeDNo1zfgscv2868mo1}
\begin{split}
\int_{x\in \T^d}|\nabla F^{a,E}_l(x+e_i/R)-\nabla F^{a,E}_l(x)|_{a}^2\leq |l|^2_{\sigma(a,E)} C_d (\frac{\|K\|_\alpha}{R^{\alpha} \lambda_{\min}(a)})^2.
 \end{split}
\end{equation}
\end{Lemma}
\begin{proof}
Observe that
\begin{equation}
\nabla .(a+E(x))\nabla\big( F^{a,E}_l(x+e_i/R)- F^{a,E}_l(x)\big)=\nabla .\Big(\big(K(x)-K(x+e_i/R)\big)\nabla F^{a,E}_l(x+e_i/R)\Big).
\end{equation}
It follows that
\begin{equation}\label{HoCoeDNo1ffzfgscv2868mo1}
\begin{split}
\int_{x\in \T^d}|\nabla F^{a,E}_l(x+e_i/R)-\nabla
F^{a,E}_l(x)|_{a}^2=& \int_{x\in \T^d}\big(\nabla
F^{a,E}_l(x+e_i/R)-\nabla
F^{a,E}_l(x)\big)\\&\big(K(x)-K(x+e_i/R)\big)\nabla
F^{a,E}_l(x+e_i/R)
\end{split}
\end{equation}
thus using Cauchy-Schwartz inequality one obtains
\begin{equation}\label{HoCoeDNo1ffzfgscv2868mo1q}
\begin{split}
\int_{x\in \T^d}|\nabla F^{a,E}_l(x+e_i/R)-\nabla F^{a,E}_l(x)|_{a}^2\leq \int_{x\in \T^d}\Big|(K(x)-K(x+e_i/R))\nabla F^{a,E}(x+e_i/R)\Big|^2_{a^{-1}}
\end{split}
\end{equation}
and the equation \ref{HoCoeDNo1zfgscv2868mo1} follows easily.
\end{proof}
It follows from lemma \ref{dslkdnlkcvnnf91112} equation \ref{HoCoeDNo1zfgscv2864} and Cauchy-Schwartz inequality that
\begin{equation}\label{HoCoeDNo1zfgscv286562}
\begin{split}
G_1\leq C_d \frac{\|K\|_\alpha}{R^{\alpha} \lambda_{\min}(a)} \|\chi^{a,P}\|_{\infty}\frac{\|a-P\|_\infty }{\lambda_{\min}(a)}\Big({^tl}\sigma_{\sym}(a, S_R P+K)l\Big)^{\frac{1}{2}}\Big({^tl}\sigma_{\sym}\big(\sigma_{\sym}(a,P),K\big)l\Big)^{\frac{1}{2}}.
 \end{split}
\end{equation}
Now we will use the following lemma which is a consequence of Stampacchia estimates \cite{St2}, \cite{St1} for elliptic operators with discontinuous coefficients (see \cite{Owh01}, appendix B, theorem B.1.1) (we remind that $\chi^{a,P}$ is uniquely defined by the cell problem and $\int_{\T^d}\chi^{a,P}(x)\,dx=0$)
\begin{Lemma}\label{ddhdvdhv777172211}
\begin{equation}\label{dkjdjdjj81222231}
\|\chi^{a,P}\|_\infty \leq C_d \big(\frac{\|a+P\|_\infty}{\lambda_{\min}(a)}\big)^{3d+2}
\end{equation}
\end{Lemma}
Thus one obtains from \eref{dkjdjdjj81222231} and \eref{HoCoeDNo1zfgscv286562} that
\begin{equation}\label{HoCoeDNo1zfgscv286562q}
\begin{split}
G_1\leq C_d \frac{\|K\|_\alpha}{R^{\alpha} \lambda_{\min}(a)} \big(\frac{\|a+P\|_\infty}{\lambda_{\min}(a)}\big)^{3d+3} \Big({^tl}\sigma_{\sym}(a, S_RP+K)l\Big)^{\frac{1}{2}}\Big({^tl}\sigma_{\sym}\big(\sigma_{\sym}(a,P),K\big)l\Big)^{\frac{1}{2}}
 \end{split}
\end{equation}
similarly,  observing that $\operatorname{div}_y \big(K(x) \nabla F^{a,E}_l(x+\frac{y}{R})\big)=0$ and integrating by part in $y$ in the equation \eref{skhsvhsjshh787711}  one obtains
\begin{equation}\label{skhsvhsjshh787711jwj71}
\begin{split}
J_4=&\sum_{i=1}^d\int_{(x,y^i)\in \T^d\times \partial^i([-\frac{1}{2},\frac{1}{2}]^d)}\big({^t\nabla F^{a,E}_l(x+(y^i+e_i)/R)}-{^t\nabla F^{a,E}_l(x+y^i/R)}\big)K(x).e_i\\& \chi^{a,P}(Rx+y^i)\nabla F^{\sigma_{\sym}(a,P),K}_l(x) \,dx\,dy^i.
\end{split}
\end{equation}
Adding equation \eref{HoCoeDNo1zfgscv2865} to equation \eref{skhsvhsjshh787711jwj71} we obtain
\begin{equation}\label{skhsvhsjshh787711jwj71000}
\begin{split}
J_4+G_2=&\sum_{i=1}^d\int_{(x,y^i)\in \T^d\times \partial^i([-\frac{1}{2},\frac{1}{2}]^d)}{^t\nabla F^{a,E}_l(x+(y^i+e_i)/R)}\big(K(x)-K(x+(y^i+e_i)/R)\big).e_i \\&\chi^{a,P}(Rx+y^i)\nabla F^{\sigma_{\sym}(a,P),K}_l(x) \,dx\,dy^i\\
&+\sum_{i=1}^d\int_{(x,y^i)\in \T^d\times \partial^i([-\frac{1}{2},\frac{1}{2}]^d)}{^t\nabla F^{a,E}_l(x+y^i/R)}\big(K(x+y^i/R)-K(x)\big).e_i\\& \chi^{a,P}(Rx+y^i)\nabla F^{\sigma_{\sym}(a,P),K}_l(x) \,dx\,dy^i
\end{split}
\end{equation}
and by Cauchy Schwartz inequality and  lemma \ref{dkjdjdjj81222231} one obtains that
\begin{equation}\label{skhsvhsjshh787711jwj710006}
\begin{split}
J_4+G_2\leq C_d \frac{\|K\|_\alpha}{R^{\alpha} \lambda_{\min}(a)} \big(\frac{\|a+P\|_\infty}{\lambda_{\min}(a)}\big)^{3d+3} \Big({^tl}\sigma_{\sym}(a, S_RP+K)l\Big)^{\frac{1}{2}}\Big({^tl}\sigma_{\sym}\big(\sigma_{\sym}(a,P),K\big)l\Big)^{\frac{1}{2}}.
\end{split}
\end{equation}
Moreover from equation \eref{dljsdbjbkjsbdkb8991} and Cauchy Schwartz inequality one easily obtains
\begin{equation}\label{dljsdbjbkjsbddddkb8991}
\begin{split}
J_3\leq C_d \frac{\|K\|_\alpha}{R^{\alpha} \lambda_{\min}(a)}  \Big({^tl}\sigma_{\sym}(a, S_RP+K)l\Big)^{\frac{1}{2}}\Big({^tl}\sigma_{\sym}\big(\sigma_{\sym}(a,P),K\big)l\Big)^{\frac{1}{2}}.
\end{split}
\end{equation}
Now we will need the following  lemma
\begin{Lemma}\label{hjgfjtffftxx1}
If $V \in (C^\infty(\T^d))^d$ is such that $\operatorname{div} V=0$ and $\int_{\T^d} V(x) dx=0$ then for $p>d$, there exists  a skew symmetric $\T^d$-periodic matrix $M$ such that $\|M\|_\infty \leq C_{d,p} \|V\|_{L^{p}(\T^d)}$ and $V=\nabla .M$.
\end{Lemma}
\begin{proof}
From the proof of lemma 4.7 of \cite{BeOw00b} one obtains that there exists a $\T^d$-periodic smooth skew-symmetric matrix $M$ such that
\begin{equation}
V_i=\sum_{j=1}^d M_{ij}
\end{equation}
and $M$ is given by
\begin{equation}
M_{ij}=B^j_i-B^i_j
\end{equation}
where $B^i_j$ are the smooth $\T^d$ periodic solutions of
\begin{equation}
\Delta B^i_j=\partial_i V_j
\end{equation}
with $0$ mean Lebesgue measure. Using the theorem 5.4 of \cite{St2} one obtains that for $p>d$
\begin{equation}
\|B^i_j\|_\infty\leq C_{d,p}\|V_j\|_{L^p(\T^d)}
\end{equation}
which proves the lemma.
\end{proof}
Let us now prove the following lemma.
\begin{Lemma}\label{dksjsdhdhdu712hs}
\begin{equation}\label{jdhsdhvcjhcvdh891}
\Big(\big(a+P(Rx+y)\big)\nabla F^{a,P}(Rx+y)-\sigma(P)\Big)_{kj}=\sum_{i=1}^d \partial_{i} M^P_{kij}(Rx+y)
\end{equation}
 where  $M^P$ is a $\T^d$ periodic $d\times d\times d$ tensor such that $M^P_{ikj}=-M^P_{kij}$ and
\begin{equation}\label{dbekjbo1}
\|M^P\|_\infty \leq C_{d,\|a+P\|_\infty/\lambda_{\min}(a)} \|a+P\|_\infty.
\end{equation}
\end{Lemma}
\begin{proof}
 From lemma \ref{hjgfjtffftxx1} one obtains that for $p>d$,
\begin{equation}
\|M^P\|_\infty \leq C_{d,p} \|\big(a+P(.))\nabla F^{a,P}(.)-\sigma(P)\big)\|_{L^{p}(\T^d)}.
\end{equation}
Using Meyers argument \cite{Mey63} one obtains that there exists $p(\|a+E\|_{\infty}/\lambda_{\min}(a))>d$ such that
\begin{equation}
\|\nabla \chi^{a,P}(.)\|_{L^{p}(\T^d)}\leq C_{d,\|a+P\|_\infty/\lambda_{\min}(a)}.
\end{equation}
Which implies equation \eref{dbekjbo1}.
\end{proof}
Using equation \eref{jdhsdhvcjhcvdh891} and integrating by part in $y$
in \eref{sajsskjbsj9sj991} one obtains
\begin{equation}\label{sahjbhsbshbsu888117221}
\begin{split}
J_2=&-\sum_{i,j,k=1}^d\int_{(x,y^i)\in \T^d\times \partial^i([-\frac{1}{2},\frac{1}{2}]^d)}\big({^t\nabla F^{a,E}_l(x+(y^i+e_i)/R)}-{^t\nabla F^{a,E}_l(x+y^i/R)}\big).e_k \\& M^P_{kij}(Rx+y^i)e_j.\nabla F^{\sigma_{\sym}(a,P),K}_l(x)\,dx\,dy^i.
\end{split}
\end{equation}
Combining this with \eref{dbekjbo1} and \eref{HoCoeDNo1zfgscv2868mo1} one obtains from Cauchy-Schwartz inequality that
\begin{equation}\label{dsldjdjndlkdsn221}
\begin{split}
J_2\leq C_{d,\|a+P\|_\infty/\lambda_{\min}(a)} \frac{\|K\|_\alpha}{R^{\alpha}\lambda_{\min}(a)}  \Big({^tl}\sigma_{\sym}(a, S_RP+K)l\Big)^{\frac{1}{2}}\Big({^tl}\sigma_{\sym}\big(\sigma_{\sym}(a,P),K\big)l\Big)^{\frac{1}{2}}.
\end{split}
\end{equation}

In conclusion we have obtained from equations \eref{sdjdhdhhdwdcdfre223}, \eref{dslnlddndklndk990091}, \eref{HoCoeDNo1zfgscv286562q},  \eref{skhsvhsjshh787711jwj710006}, \eref{dljsdbjbkjsbddddkb8991} and \eref{dsldjdjndlkdsn221} that
\begin{equation}\label{sdjdhdhhdwdcdfre2ss3123}
\begin{split}
\bigg[ \Big({^tl}\sigma_{\sym}(a, S_R P+K)l\Big)^{\frac{1}{2}}-&\Big({^tl}\sigma_{\sym}\big(\sigma_{\sym}(a,P),K\big)l\Big)^{\frac{1}{2}}\bigg]^2\leq C_{d,\|a+P\|_\infty/\lambda_{\min}(a)} \frac{\|K\|_\alpha}{R^{\alpha}\lambda_{\min}(a)}  \\&\Big({^tl}\sigma_{\sym}(a, S_RP+K)l\Big)^{\frac{1}{2}}\Big({^tl}\sigma_{\sym}\big(\sigma_{\sym}(a,P),K\big)l\Big)^{\frac{1}{2}}.
\end{split}
\end{equation}
Now we will use the following lemma whose proof is trivial
\begin{Lemma}\label{sdhdhdbdhbb7711221aa}
If $(X-Y)^2\leq \delta XY$ then $X/Y \leq (1+8\sqrt{\delta})^2$
\end{Lemma}
And the estimate \eref{dkjdjbj8981111} is a simple consequence of \eref{sdjdhdhhdwdcdfre2ss3123} and lemma \ref{sdhdhdbdhbb7711221aa} which proves the theorem.

\subsection{Averaging with $n$ scales: Proof of  theorem \ref{asjhsskbjs801}}\label{sksjkjbkjcbc987191}
The proof of theorem \ref{asjhsskbjs801} is based on theorem \ref{0kjhhbxdhbo1222} and a reverse induction. It is important to note that contrary to reiterated homogenization, here the larger scales are homogenized first, this reversion in the inductive process is essential to obtain sharp estimates.
Observe that by the variational formula \ref{HoCoeDNorPre12} one has for $\gamma>0$, $B\in \M_{d,\sym}$ and $K\in \SL(\T^d)$,
\begin{equation}\label{ddbxcjbc983921}
\sigma_{sym}(B,\gamma K)=\gamma \sigma_{sym}(\frac{B}{\gamma}, K).
\end{equation}
From this we deduce that for $p\in \{0,\ldots,n-1\}$
\begin{equation}\label{dhdhdhdhdhhjhh81121}
\sigma_{sym}(\sigma_{sym}(B,E^p),\frac{1}{\gamma_p}\Gamma^{p+1,n})=\frac{\gamma_{p+1}}{\gamma_p}\sigma_{sym}(\frac{\gamma_p}{\gamma_{p+1}}\sigma_{sym}(B,E^p),\frac{1}{\gamma_{p+1}}\Gamma^{p+1,n}).
\end{equation}
Combining this with the theorem \ref{0kjhhbxdhbo1222} one obtains that for $p\in \{0,\ldots,n-1\}$
\begin{equation}\label{ddddddndjun8811221}
\begin{split}
\sigma_{sym}(B,\frac{1}{\gamma_p}\Gamma^{p,n})\leq \frac{\gamma_{p+1}}{\gamma_p}\sigma_{sym}(\frac{\gamma_p}{\gamma_{p+1}}\sigma_{sym}(B,E^p),\frac{1}{\gamma_{p+1}}\Gamma^{p+1,n}) (1+\epsilon_{p+1}(B))^4
\end{split}
\end{equation}
\begin{equation}\label{dlijicjijij90n21}
\begin{split}
\sigma_{sym}(B,\frac{1}{\gamma_p}\Gamma^{p,n})\geq \frac{\gamma_{p+1}}{\gamma_p}\sigma_{sym}(\frac{\gamma_p}{\gamma_{p+1}}\sigma_{sym}(B,E^p),\frac{1}{\gamma_{p+1}}\Gamma^{p+1,n}) (1+\epsilon_{p+1}(B))^{-4}
\end{split}
\end{equation}
with
\begin{equation}
\begin{split}
\epsilon_p(B)=& \Big(\frac{\gamma_{p}K_\alpha}{\gamma_{p-1}r_p^\alpha \lambda_{\min}(B) (1-\gamma_{\max}/\rho_{\min}^\alpha)}\Big)^\frac{1}{2}f\big((d,(\lambda_{\max}(B)+K_0)/\lambda_{\min}(B)\big)).
\end{split}
\end{equation}
Then one obtains by a simple induction that
\begin{equation}
\begin{split}
\gamma_{n+1} A^{n+1} \prod_{p=0}^{n-1}(1+\epsilon_{p+1}(A^{p}))^{-4} \leq \sigma_{sym}(a,\Gamma^{0,n})\leq \gamma_{n+1} A^{n+1} \prod_{p=0}^{n-1}(1+\epsilon_{p+1}(A^{p}))^4.
\end{split}
\end{equation}
Where $A^{k}$, is the \emph{renormalization coreization sequence} given in  definition \ref{sdjhdsgdsuzdggggg2876261212} which proves theorem \ref{asjhsskbjs801}.

\subsection{Diagnosis of \emph{renormalization core}'s pathologies: proofs}\label{sdjjddgjsdgd662fgs52121}
Let $a\in \M_{d,\sym}$ and $E\in \SL(\T^d)$, it is well known  (\cite{AvMa91}) and a simple consequence of \eref{HoCoeDNorPre12} and \eref{HoCoeDNorPre14} that
\begin{equation}\label{Jshkshjbks871}
a\leq \sigma_{\sym}(a,E)\leq a+ \int_{\T^d}{^tE(x)a^{-1}E(x)}\,dx.
\end{equation}
Then the following proposition follows from \eref{wdkbkwjdebkljb81}, \eref{Jshkshjbks871} and a simple induction on the number of scales.
\begin{Proposition}\label{dsdnlnllk9111}
For all $n\in \N$
\begin{equation}
\gamma_n^{-1} I_d\leq A^n \leq (\kappa/\gamma_{n}) I_d+\sum_{p=0}^{n-1} (\gamma_p/\gamma_n)\int_{T^d_1}{^tE^p(x)(A^p)^{-1}E^p(x)}\,dx.
\end{equation}
\end{Proposition}
Theorems \ref{djshddsadsdbhdaaaab881} and \ref{djshddaqadsdbhdb881} are straightforward consequences of proposition \ref{dsdnlnllk9111}.
We will need the following proposition giving isotropic estimates on anisotropic viscosities.
\begin{Proposition}\label{skjsbkjbjbdjb89191}
For  $a\in \M_{d,\sym}$ and $E\in \SL(\T^d)$, one has for all $l\in (\R^d)^*$
\begin{equation}
\begin{split}
\big(\frac{\lambda_{\max}(a)}{\lambda_{\min}(a)}\big)^{-\frac{1}{2}} \leq \frac{^tl\sigma_{\sym}(a,E)l}{^tl\sigma_{\sym}\Big(\big(\lambda_{\max}(a)\lambda_{\min}(a)\big)^\frac{1}{2} I_d,E\Big)l
}\leq \big(\frac{\lambda_{\max}(a)}{\lambda_{\min}(a)}\big)^{\frac{1}{2}}.
\end{split}
\end{equation}
\end{Proposition}
\begin{proof}
By the variational formula \ref{HoCoeDNorPre12} one has for $\xi\in \R^d$
\begin{equation}
\begin{split}
|\xi|_{\sigma_{\sym}^{-1}(a,E)}^2=&\inf_{(f,H)\in C^\infty(\T^d)\times \S(\T^d)}\int_{\T^d}|\xi-\nabla .H+ E\nabla f|_{a^{-1}}^2\,dx + \int_{\T^d}|\nabla f(x)|_a^2\,dx\\
\leq & \big(\lambda_{\max}(a)/\lambda_{\min}(a)\big)^\frac{1}{2} \inf_{(f,H)\in C^\infty(\T^d)\times
\S(\T^d)}\big(\lambda_{\max}(a)\lambda_{\min}(a)\big)^{-\frac{1}{2}}\int_{\T^d}|\xi-\nabla .H+ E\nabla
f|^2\,dx\\& + \big(\lambda_{\max}(a)\lambda_{\min}(a)\big)^\frac{1}{2}\int_{\T^d}|\nabla f(x)|^2\,dx.
\end{split}
\end{equation}
It follows that
\begin{equation}
\begin{split}
\sigma_{\sym}(a,E)\geq \big(\lambda_{\max}(a)/\lambda_{\min}(a)\big)^{-\frac{1}{2}} \sigma_{\sym}\Big(\big(\lambda_{\max}(a)\lambda_{\min}(a)\big)^\frac{1}{2} I_d,E\Big).
\end{split}
\end{equation}
Similarly from the variational formula \ref{HoCoeDNorPre14} one obtains that for $l\in \R^d$
\begin{equation}
\begin{split}
|l|_{\sigma_{\sym}}^2=&\inf_{\xi \perp l, (f,H) \in C^\infty(\T^d)\times \S(\T^d)}\int_{\T^d}|\xi-\nabla .H-E(l-\nabla f)|^2_{a^{-1}}\,dx+\int_{\T^d}|l-\nabla f|^2_{a}\,dx\\
\leq & \big(\lambda_{\max}(a)/\lambda_{\min}(a)\big)^\frac{1}{2}\inf_{\xi \perp l, (f,H) \in
C^\infty(\T^d)\times \S(\T^d)}\\& \big(\lambda_{\max}(a)\lambda_{\min}(a)\big)^{-\frac{1}{2}}
\int_{\T^d}|\xi-\nabla .H-E(l-\nabla f)|^2\,dx +\big(\lambda_{\max}(a)\lambda_{\min}(a)\big)^\frac{1}{2}
\int_{\T^d}|l-\nabla f|^2_{a}\,dx.
\end{split}
\end{equation}
Which leads us to
\begin{equation}
\begin{split}
\sigma_{\sym}(a,E)\leq \big(\lambda_{\max}(a)/\lambda_{\min}(a)\big)^{\frac{1}{2}} \sigma_{\sym}\Big(\big(\lambda_{\max}(a)\lambda_{\min}(a)\big)^\frac{1}{2} I_d,E\Big).
\end{split}
\end{equation}
\end{proof}
A direct consequence of proposition \ref{skjsbkjbjbdjb89191} is the following corollary which controls the
minimal and maximal enhancement of the conductivity in the flow associated to the stream matrix $E$ by the
geometric mean $\big(\lambda_{\max}(a)\lambda_{\min}(a)\big)^\frac{1}{2}$ of the maximal and minimal eigenvalues
of $a$.
\begin{Corollary}\label{ekjdbkjbkjbf89811}
\begin{equation}
\begin{split}
\frac{\lambda_{\min}\big( \sigma_{\sym}(a,E)\big)}{\lambda_{\min}(a)}\geq \frac{\lambda_{\min}\bigg(\sigma_{\sym}\Big(\big(\lambda_{\max}(a)\lambda_{\min}(a)\big)^\frac{1}{2} I_d,E\Big)\bigg)}{\big(\lambda_{\max}(a)\lambda_{\min}(a)\big)^\frac{1}{2}}
\end{split}
\end{equation}
\begin{equation}
\begin{split}
\frac{\lambda_{\max}\big( \sigma_{\sym}(a,E)\big)}{\lambda_{\max}(a)}\leq \frac{\lambda_{\max}\bigg(\sigma_{\sym}\Big(\big(\lambda_{\max}(a)\lambda_{\min}(a)\big)^\frac{1}{2} I_d,E\Big)\bigg)}{\big(\lambda_{\max}(a)\lambda_{\min}(a)\big)^\frac{1}{2}}.
\end{split}
\end{equation}
\end{Corollary}
It is then a simple consequence of corollary \ref{ekjdbkjbkjbf89811} that
\begin{Proposition}\label{ssjhbkdbjdb899111}
\begin{equation}
\frac{\lambda_{\min}(\sigma_{\sym}(a,E))}{\lambda_{\min}(a)}\geq V\Big(\big(\lambda_{\min}(a) \lambda_{\max}(a)\big)^\frac{1}{2},E\Big)
\end{equation}
and
\begin{equation}
\frac{\lambda_{\max}(\sigma_{\sym}(a,E))}{\lambda_{\max}(a)}\leq W\Big(\big(\lambda_{\min}(a) \lambda_{\max}(a)\big)^\frac{1}{2},E\Big).
\end{equation}
\end{Proposition}
From  proposition \ref{ssjhbkdbjdb899111} one obtains that for $n\in \N$,
\begin{equation}\label{ddkdjbkdjbdjkdbdjd989871}
\frac{\lambda_{\min}(A^{n+1})}{\lambda_{\min}(A^n)}\geq \frac{\gamma_n}{\gamma_{n+1}} V\Big(\big(\lambda_{\min}(A^n)\lambda_{\max}(A^n)\big)^\frac{1}{2},E^n\Big).
\end{equation}
It follows from the equation \eref{ddkdjbkdjbdjkdbdjd989871} and the monotony of $V$ that
\begin{equation}\label{ddkdjbkdjbdjkdbdjd989871q}
\frac{\lambda_{\min}(A^{n+1})}{\lambda_{\min}(A^n)}\geq \frac{1}{\gamma_{max}} V\big(\lambda_{\min}(A^n) (\frac{\lambda_{\max}(A^n)}{\lambda_{\min}(A^n)})^\frac{1}{2}\big)
\end{equation}
 it follows from \eref{ddkdjbkdjbdjkdbdjd989871q} that
 $\lambda_{\min}(A^n)$ is increasing if it belongs to $\Big(0, (\frac{\lambda_{\max}(A^n)}{\lambda_{\min}(A^n)})^{-\frac{1}{2}} V^{-1}(\gamma_{\max})\Big)$; which implies equation \eref{ddkdjbkdjbdjkdbdjd98987101} of theorem \ref{djshddsadsdbhdaaaab881} and equation \eref{ddkdjbkdjbdjkdbdjad98987101} of theorem \ref{sudjzdd78727hg712}.\\
Now,
observe that from the variational formulation \eref{sshbskjhbcbb98811} one obtains that
\begin{equation}\label{sshbskjhbcbb98812}
W(\zeta, E)\leq 1+\zeta^{-2}\lambda_{\max}\Big(\int_{\T^d}{^tE(x) E(x)}dx\Big).
\end{equation}
It follows from the proposition  \ref{ssjhbkdbjdb899111} that
\begin{equation}
\frac{\lambda_{\max}(\sigma_{\sym}(a,E))}{\lambda_{\max}(a)}\leq 1+C_d K_0^2 \big(\lambda_{\min}(a) \lambda_{\max}(a)\big)^{-1}.
\end{equation}
Thus  one obtains for all $n\in \N$
\begin{equation}\label{ssjhvjhxvhsjvx8771s}
\frac{\lambda_{\max}(A^{n+1})}{\lambda_{\max}(A^n)}\leq \gamma_{\min}^{-1}\Big(1+C_d K_0^2 \frac{\lambda_{\max}(A^n)}{\lambda_{\min}(A^n)} \big( \lambda_{\max}(A^n)\big)^{-2} \Big).
\end{equation}
It follows from \eref{ssjhvjhxvhsjvx8771s} that $\lambda_{\max}(A^n)$ is decreasing if it belongs to
$\Big( \big(C_d K_0^2 \frac{\lambda_{\max}(A^n)}{\lambda_{\min}(A^n)} (\gamma_{\min}-1)^{-1}\big)^\frac{1}{2},\infty\Big)$
; which implies the equation \eref{ddkdjbkdjbdjkdbdjd98987101q} of theorem \ref{djshddsadsdbhdaaaab881} and equation \eref{ddkdjbkdjbdjkdbsdjd98987101q} of theorem \ref{dkdccizueztU721yys}.\\
Now observe that by proposition \ref{ssjhbkdbjdb899111} one has
\begin{equation}\label{ddkdjbkdjbdjkdbdjd9aa89871q}
\frac{\lambda_{\max}(A^{n+1})}{\lambda_{\max}(A^n)}\leq \frac{1}{\gamma_{min}} W\big(\lambda_{\min}(A^n) (\frac{\lambda_{\max}(A^n)}{\lambda_{\min}(A^n)})^\frac{1}{2}\big)
\end{equation}
which proves theorem \ref{djsswahdddsdbhdaaaab881} since $W$ is decreasing.\\
Now if the flow is self-similar and isotropic, theorem \ref{sajssgszsgzs8712} is a simple consequence of the following recursive relation:
\begin{equation}\label{ddkdjbkdjbdjkdswsbdjd989871q}
\frac{\lambda(A^{n+1})}{\lambda(A^n)}= \frac{1}{\gamma} V\big(\lambda(A^n)\big).
\end{equation}

\subsection{Super diffusion: proofs}\label{eefifgirfgzrg898731we}
\subsubsection{A variational formula for the exit times}\label{djhsdgkdjgfdhdgsfdtsf443}
Let $\Omega$ be a smooth subset of $\R^d$, we write for $a\in \M_{d,\sym}$ and $E$ a  skew symmetric matrix with coefficients in  $L^\infty(\bar{\Omega})$,
\begin{equation}
\psi^{a,E}=\E_x[\tau^{a,E}(\Omega)]
\end{equation}
the expectation of the exit time from $\Omega$ of the diffusion associated to the generator $\nabla .(a+E)\nabla$ started from $x$. Observe that $\psi^{a,E}$ can be defined as the weak solution of the following equation with null Dirichlet boundary condition on $\partial \Omega$,
\begin{equation}\label{erjbjdfbjfbfb}
\nabla .\Big( \big(a+E(x)\big)\nabla\psi^{a,E}(x)\Big)=-1.
\end{equation}
We will need the following variational formulation for the mean exit times.

\begin{Theorem}\label{xsddhdxhxvhxv898812}
\begin{equation}\label{sjdhghdgd6762233}
\int_{\Omega}\E_x[\tau^{a,E}(\Omega)] \,dx=\sup_{f\in C^\infty_0(\Omega),H\in \S(\bar{\Omega})}\Big[  2\int_{\Omega}f(x)dx
-\int_{\Omega}|\nabla f|_{a}^2 dx-\int_{\Omega}|\nabla .H +E\nabla f|_{a^{-1}}^2 dx\Big]
\end{equation}
\end{Theorem}
Where the minimization \eref{sjdhghdgd6762233} is done over   smooth functions $f$ on $\Omega$, null on $\partial \Omega$ and  smooth skew symmetric matrices $H$  on $\bar{\Omega}$.
From theorem \ref{xsddhdxhxvhxv898812} we deduce the following corollary
\begin{Corollary}\label{xsddhdxhxvhxv883398812}
\begin{equation}
\int_{\Omega}\E_x[\tau^{a+\lambda I_d,0}(\Omega)] \,dx \leq \int_{\Omega}\E_x[\tau^{a,E}(\Omega)] \,dx\leq \int_{\Omega}\E_x[\tau^{a,0}(\Omega)] \,dx
\end{equation}
with
\begin{equation}
\lambda:=\sup_{x\in \Omega}\lambda_{\max}\big({^tE(x)a^{-1}E(x)}\big)
\end{equation}
\end{Corollary}
Let us now prove  theorem \ref{xsddhdxhxvhxv898812}.
By density we can first assume $E$ to be smooth. Our purpose is to show that
\begin{equation}
\int_{\Omega}\E_x[\tau^{a,E}(\Omega)] \,dx=-2\inf_{f\in C^\infty_0(\Omega),H\in \S(\bar{\Omega})}\Big[\frac{1}{2}\int_{\Omega}|\nabla .H +(a-E)\nabla f|_{a^{-1}}^2 dx - \int_{\Omega}f(x)dx\Big].
\end{equation}
 By considering variations around the minimum one obtains that
\begin{equation}
\nabla .H_0 +(a-E)\nabla f_0=a \nabla \psi(x)
\end{equation}
with $\psi=0$ on $\partial \Omega$ and
\begin{equation}
\nabla .(a+E) \nabla \psi(x)=-1.
\end{equation}
From which one obtains that $\psi(x)=\E_x[\tau^{E}(\Omega)]$ and $f_0(x)=\big(\E_x[\tau^{E}(\Omega)]+\E_x[\tau^{-E}(\Omega)]\big)/2$. Thus at the minimum
\begin{equation}
\begin{split}
-\inf_{H,f}\Big[\frac{1}{2}\int_{\Omega}|\nabla .H +(a-E)\nabla f|_{a^{-1}}^2 dx - \int_{\Omega}f(x)dx\Big]&=
-\Big[\frac{1}{2}\int_{\Omega}{^t \nabla f_0}(a+E)\nabla \psi dx - \int_{\Omega}f_0(x)dx\Big]\\&=-1/2  \int_{\Omega}f_0(x)dx
\end{split}
\end{equation}
since
\begin{equation}
\begin{split}
\int_{\Omega}{^t \nabla f_0}(a+E)\nabla \psi(x) dx=\int_{\Omega}|a \nabla \psi(x)|^2_{a^{-1}} dx
\end{split}
\end{equation}
but also
 \begin{equation}
\begin{split}
\int_{\Omega}|a \nabla \psi(x)|^2_{a^{-1}} dx=\int_{\Omega} {^t \nabla\psi(x)}(a+E)\nabla \psi(x) dx=\int_{\Omega}\psi(x) dx
\end{split}
\end{equation}
which leads to the result, which can be written as \eref{sjdhghdgd6762233}.

\subsubsection{Averaging with two scales the exit times}
We will use the notation of subsection \ref{djhsdgkdjgfdhdgsfdtsf443} and assume that
\begin{equation}
E=P(Rx)+K(x).
\end{equation}
Where $x\in\Omega$, $R\in [2,\infty)$, $P$ belongs to $\SL(\T^d)$ and $K$ is a Lipschitz-continuous skew symmetric matrix on $\R^d$ ($\alpha=1$). Our purpose is to obtain sharp quantitative estimates on the mean exit time.
\begin{equation}\label{erjbjdfbjf3333bfb}
\int_{\Omega}\psi^{a,E}(x)\,dx.
\end{equation}
It follows from theorem \ref{xsddhdxhxvhxv898812} that the mean exit time \eref{erjbjdfbjf3333bfb} is continuous in $L^\infty$ norm with respect to $E$, thus we can by density assume $E,P$ and $K$ to be smooth and $\psi^{a,E}$ shall be a strong solution of \eref{erjbjdfbjfbfb}.\\
To estimate \eref{erjbjdfbjf3333bfb} we will need to introduce a relative translation with respect to the fast scale associated to the medium $E$, i.e. we introduce for $x,y\in \Omega\times [0,1]^d$, $E(x,y)$ as
\begin{equation}
E(x,y):=P(Rx+y)+K(x).
\end{equation}
We will write for $y\in [0,1]^d$, $\psi^{a,E}(x,y)$ the strong solution of the following equation with null Dirichlet boundary condition on $\partial \Omega$.
\begin{equation}\label{erjbjdfbjfbfbq}
\nabla_x\Big( \big(a+E(x,y)\big)\nabla_x\psi^{a,E}(x,y)\Big)=-1
\end{equation}
Let us define
\begin{equation}\label{dssjhcchc6}
\begin{split}
J:=&-\int_{x\in \Omega,y\in [0,1]^d}{^t\nabla \psi^{a,E}}(x,y)\big((a+P(Rx+y))\nabla F^{a,P}(Rx+y)-\sigma(a,P)\big) \nabla \psi^{\sigma_{\sym}(a,P),K}(x)\,dx\,dy\\
&
+\int_{x\in \Omega,y\in [0,1]^d}{^t\nabla \psi^{a,E}}(x,y)\big(a-P(Rx+y)\big)\nabla \chi^{a,P}(Rx+y)\nabla \psi^{\sigma_{\sym}(a,P),K}(x)\,dx\,dy.
\end{split}
\end{equation}
Now we will show that $J$ controls the multi-scale homogenization associated to $\psi(x,y)$
\begin{Proposition}\label{duudgddg7782727717221}
One has
\begin{equation}\label{dssjhcchssssc6}
\begin{split}
\Big(\big(\int_{x\in \Omega,y\in [0,1]^d} \psi^{a,E}(x,y)\,dx\,dy\big)^\frac{1}{2}-\big(\int_{x\in \Omega}\psi^{\sigma_{\sym}(a,P),K}(x)\,dx\big)^\frac{1}{2}\Big)^2\leq J.
\end{split}
\end{equation}
\end{Proposition}
\begin{proof}
Let us write
\begin{equation}\label{dssjhcchc50}
I=\int_{x\in \Omega,y\in [0,1]^d}{^t\nabla \psi^{a,E}}(x,y) a\nabla F^{a,P}(Rx+y)\nabla \psi^{\sigma_{\sym}(a,P),K}(x)\,dx\,dy.
\end{equation}
Observing that
\begin{equation}\label{wdjbejebejbjb8112}
\begin{split}
\int_{x\in \Omega,y\in [0,1]^d}{^t\nabla \psi^{a,E}}(x,y) a\nabla \psi^{a,E}\,dx\,dy&=\int_{x\in \Omega,y\in [0,1]^d}{^t\nabla \psi^{a,E}}(x,y) \big(a+E(x,y)\big)\nabla \psi^{a,E}\,dx\,dy\\
&=\int_{x\in \Omega,y\in [0,1]^d} \psi^{a,E}(x,y)\,dx\,dy
\end{split}
\end{equation}
and
\begin{equation}
\begin{split}
\int_{x\in \Omega,y\in [0,1]^d}{^t\nabla \psi^{\sigma_{\sym}(a,P),K}}(x){^t\nabla F^{a,P}(Rx+y)} &a\nabla F^{a,P}(Rx+y)\nabla \psi^{\sigma_{\sym}(a,P),K}(x)\,dx\,dy\\&=\int_{x\in \Omega}\psi^{\sigma_{\sym}(a,P),K}(x)\,dx
\end{split}
\end{equation}
one obtains by Cauchy-Schwartz inequality from \eref{dssjhcchc50} that
\begin{equation}\label{dhsjdhddh98811}
I\leq\Big(\int_{x\in \Omega,y\in [0,1]^d} \psi^{a,E}(x,y)\,dx\,dy\int_{x\in \Omega}\psi^{\sigma_{\sym}(a,P),K}(x)\,dx\Big)^\frac{1}{2}.
\end{equation}
Now let us polarize $I$ as
\begin{equation}\label{dssjhcchc}
I=\big(I_1+I_2\big)/2
\end{equation}
with
\begin{equation}\label{dssjhcchc1}
I_1=\int_{x\in \Omega,y\in [0,1]^d}{^t\nabla \psi^{a,E}}(x,y)\big(a-E(x,y)\big)\nabla F^{a,P}(Rx+y)\nabla \psi^{\sigma_{\sym}(a,P),K}(x)\,dx\,dy
\end{equation}
and
\begin{equation}\label{dssjhcchc2}
I_2=\int_{x\in \Omega,y\in [0,1]^d}{^t\nabla \psi^{a,E}}(x,y)\big(a+E(x,y)\big)\nabla F^{a,P}(Rx+y)\nabla \psi^{\sigma_{\sym}(a,P),K}(x)\,dx\,dy.
\end{equation}
Using $\nabla .\big(a+E(x+y/R)\big){\nabla \psi^{a,E}}(x,y)=-1$ one obtains that
\begin{equation}\label{dssjhcchc3}
\begin{split}
I_1=&\int_{x\in \Omega}\psi^{\sigma_{\sym}(a,P),K}(x)\,dx\\
&-
\int_{x\in \Omega,y\in [0,1]^d}{^t\nabla \psi^{a,E}}(x,y)\big(a-E(x,y)\big)\nabla \chi^{a,P}(Rx+y)\nabla \psi^{\sigma_{\sym}(a,P),K}(x)\,dx\,dy.
\end{split}
\end{equation}
Moreover
\begin{equation}\label{dssjhcchc4}
\begin{split}
I_2=&\int_{x\in \Omega,y\in [0,1]^d}{^t\nabla \psi^{a,E}}(x,y)\big(\sigma(a,P)+K(x)\big) \nabla \psi^{\sigma_{\sym}(a,P),K}(x)\,dx\,dy\\
&+\int_{x\in \Omega,y\in [0,1]^d}{^t\nabla \psi^{a,E}}(x,y)\big((a+P(Rx+y))\nabla F^{a,P}(Rx+y)-\sigma(a,P)\big) \nabla \psi^{\sigma_{\sym}(a,P),K}(x)\,dx\,dy\\
&-\int_{x\in \Omega,y\in [0,1]^d}{^t\nabla \psi^{a,E}}(x,y)K(x)\nabla \chi^{a,P} (Rx+y)\nabla \psi^{\sigma_{\sym}(a,P),K}(x)\,dx\,dy.
\end{split}
\end{equation}
And observing that
\begin{equation}
\begin{split}
\int_{x\in \Omega,y\in [0,1]^d}{^t\nabla \psi^{a,E}}(x,y)\big(\sigma(a,P)+K(x)\big) \nabla \psi^{\sigma_{\sym}(a,P),K}(x)\,dx\,dy=\int_{x\in \Omega,y\in [0,1]^d}\psi^{a,E}(x,y)\,dx\,dy
\end{split}
\end{equation}
one obtains from the combination of \eref{dssjhcchc}, \eref{dssjhcchc3} and \eref{dssjhcchc4} that
\begin{equation}\label{dssjhcchc5}
2 I=\int_{x\in \Omega,y\in [0,1]^d}\psi^{a,E}(x,y)\,dx\,dy+\int_{x\in \Omega}\psi^{\sigma_{\sym}(a,P),K}(x)\,dx-J
\end{equation}
with $J$ given by equation \eref{dssjhcchc6}. Next one easily obtains \eref{dssjhcchssssc6}
 from \eref{dssjhcchc5} and \eref{dhsjdhddh98811}.
\end{proof}
We will now show that $J$ acts as an error term.
We will need the following lemmas.
\begin{Lemma}\label{sdjhddhsgddgg7812}
Let $\sigma$ be a positive definite symmetric constant matrix.
There exists a constant $C_{d}$ depending only on the dimension $d$ such that for any function $f\in C^2_0(\Omega)$
one has
\begin{equation}\label{dskdjdjdhdh88112}
\sum_{i,j=1}^d\int_{\Omega} \big(\partial_i\partial_j f(x)\big)^2\,dx\leq C_{d} \big(\lambda_{\min}(\sigma)\big)^{-2}
\int_{\Omega} \big(\nabla \sigma \nabla f(x)\big)^2\,dx.
\end{equation}
\end{Lemma}
\begin{proof}
When $\Omega=\R^d$ and $f\in C^\infty_0(\R^2)$, the inequality  \eref{dskdjdjdhdh88112} is  standard,
 we refer to the theorem 1.7 of \cite{Si72}. When $\Omega$ is a bounded open subset of $\R^d$ with smooth boundary
the proof follows trivially from the density of $C^\infty_0(\R^2)$ in $C^2_0(\Omega)$.
\end{proof}
We write $T(\Omega)$ the set of smooth $d$-dimensional vector field on $\bar{\Omega}$,
 $\xi \in \big(C^\infty(\bar{\Omega})\big)^d$ such that
\begin{equation}\label{shsjahgssszgzs676211}
\forall z\in \partial \Omega,\quad \xi(z).n(z)=1
\end{equation}
 where  $\partial \Omega$ is the boundary of $\Omega$ and $n(z)$ the exterior orthonormal vector at the point
$z$ of the boundary.
For $\Omega$ a bounded open subset of $\R^d$ with smooth boundary we write $\Theta(\Omega)$ the following
isoperimetric constant associated to $\Omega$
\begin{equation}
\Theta(\Omega):=\inf_{\xi \in T(\Omega)}\max\big(\|\xi\|_\infty,\|\nabla \xi\|_\infty\big).
\end{equation}
\begin{Lemma}\label{sdjdgjhddgg787112}
We have
\begin{equation}\label{dsqayaaaydaqaaadyyaaydjkj1s411c6aaassab}
\begin{split}
\int_{z\in \partial \Omega,y\in [0,1]^d} & \big({^t\nabla} \psi^{a,E}(z,y)a \nabla
\psi^{\sigma_{\sym}(a,P),K}(z)\big)  \,dz\,dy \leq C_{d,\|a+P\|_\infty/\lambda_{\min}(a)}\Theta(\Omega)
 \\&
\Bigg(\big(1+ \frac{\|K\|_{1}}{\lambda_{\min}(a)} \big)
\Big(\int_{x\in \Omega,y\in [0,1]^d} \psi^{a,E}(x,y)\,dx\,dy
\int_{x\in \Omega}\psi^{\sigma_{\sym}(a,P),K}(x)\,dx\Big)^\frac{1}{2}
\\&+
\Big(\frac{\Vol(\Omega)}{\lambda_{\min}(a)}\Big)^\frac{1}{2}
\Big(\int_{x\in \Omega}\psi^{\sigma_{\sym}(a,P),K}(x)\,dx
+\int_{x\in \Omega,y\in [0,1]^d} \psi^{a,E}(x,y)\,dx\,dy\Big)^\frac{1}{2}\Bigg).
\end{split}
\end{equation}
\end{Lemma}
\begin{proof}
Let $f$ and $v$ be a smooth function and a smooth vector field on $\bar{\Omega}$
 we will  use the following Green formula
\begin{equation}\label{sdkddkdjdduu8822}
\int_{\Omega}f(x)\diiv v(x)\,dx=-\int_{\Omega}\nabla f(x) v(x)\,dx+\int_{\partial \Omega} f(z) \big(v(z).n(z)\big)\,dz.
\end{equation}
Where  $dz$ is the measure surface at the boundary. Let $\xi \in T(\Omega)$. Let us write
\begin{equation}\label{dssjssaashcchdddjkj1411c6}
\begin{split}
G=\sum_{i,j,k=1}^d \int_{x\in \Omega,y\in [0,1]^d} \partial_i \psi^{a,E}(x,y) \big(a-P(Rx+y)\big)_{ij}\partial_j \xi_k(x) \partial_k \psi^{\sigma_{\sym}(a,P),K}(x)\,dx\,dy.
\end{split}
\end{equation}
Applying formula \eref{sdkddkdjdduu8822} to equation \eref{dssjssaashcchdddjkj1411c6}  with
$\nabla f=\nabla \xi_k(x)$
we obtain that
\begin{equation}\label{dssjhccssshdddjkj1411c6aaaa}
\begin{split}
G=G_1+G_2+G_3
\end{split}
\end{equation}
with (using the skew symmetry of $P_{ij}$ in $ij$)
\begin{equation}\label{dssjhcchaadddjkj1411c6aaaab}
\begin{split}
G_1=-\sum_{i,j,k=1}^d \int_{x\in \Omega,y\in [0,1]^d} \partial_j \Big(\big(  a+P(Rx+y)\big)_{ji}\partial_i \psi^{a,E}(x,y)\Big) \xi_k(x)\partial_k \psi^{\sigma_{\sym}(a,P),K}(x)\,dx\,dy
\end{split}
\end{equation}
\begin{equation}\label{daassjhcchdddjkj1s411c6aaassab}
\begin{split}
G_2=&-\sum_{i,j,k=1}^d \int_{x\in \Omega,y\in [0,1]^d}  \partial_i \psi^{a,E}(x,y)\big(a-P(Rx+y)\big)_{ij} \xi_k(x) \\&\partial_j\partial_k \psi^{\sigma_{\sym}(a,P),K}(x)\,dx\,dy
\end{split}
\end{equation}
\begin{equation}\label{dsaysjhcchddssadjkj1s411c6aaassab}
\begin{split}
G_3=&\sum_{i,j,k=1}^d \int_{z\in \partial \Omega,y\in [0,1]^d}  \partial_i \psi^{a,E}(z,y)\big(a-P(Rz+y)\big)_{ij} \xi_k(z)\\&n_{j}(z)\partial_k \psi^{\sigma_{\sym}(a,P),K}(z) \,\,dz\,dy
\end{split}
\end{equation}
Where $n_j$ are the coordinates of the exterior orthonormal vector $n(z)$. Using the fact that $\nabla \psi^{a,E}(z,y)$
and $\nabla \psi^{\sigma_{\sym}(a,P),K}(z)$ are parallel to $n(z)$ at the boundary of $\Omega$ and both heading towards the opposite direction of $n$, we obtain that (using the skew symmetry of $P_{ij}$ in $ij$)
\begin{equation}\label{dsaysjhcchyyddyyydjkj1s411c6aaassab}
\begin{split}
G_3=& \int_{z\in \partial \Omega,y\in [0,1]^d}  \big({^t\nabla} \psi^{a,E}(z,y)a \nabla \psi^{\sigma_{\sym}(a,P),K}(z)\big)\big(\xi(z).n(z)\big)  \,dz\,dy.
\end{split}
\end{equation}
Thus by equation \eref{shsjahgssszgzs676211},
\begin{equation}\label{dsaysjhcchyydaaadyyydjkj1s411c6aaassab}
\begin{split}
G_3=& \int_{z\in \partial \Omega,y\in [0,1]^d}  \big({^t\nabla} \psi^{a,E}(z,y)a \nabla \psi^{\sigma_{\sym}(a,P),K}(z)\big)  \,dz\,dy.
\end{split}
\end{equation}
Now, by Cauchy-Schwartz inequality we obtain from \eref{dssjssaashcchdddjkj1411c6}
\begin{equation}\label{dssjssaashcchdddjkj1411c6ac}
\begin{split}
|G|\leq C_{d}\|a+P\|_{\infty}(\lambda_{\min}(a))^{-1}\|\nabla \xi\|_\infty
\Big(\int_{x\in \Omega,y\in [0,1]^d} \psi^{a,E}(x,y)\,dx\,dy\int_{x\in \Omega}\psi^{\sigma_{\sym}(a,P),K}(x)\,dx\Big)^\frac{1}{2}.
\end{split}
\end{equation}
Using Cauchy-Schwartz inequality and $\nabla .(a+P(Rx+y))\nabla\psi^{a,E}(x,y)=-1-\nabla .K(x)\nabla\psi^{a,E}(x,y)$ we obtain
from equation
\eref{dssjhcchaadddjkj1411c6aaaab} that
\begin{equation}\label{dssjhcchaadddjkj1411c6aaaabe}
\begin{split}
|G_1|\leq &C_{d}\big(\lambda_{\min}(\sigma_{\sym}(a,P))\big)^{-\frac{1}{2}}\| \xi\|_\infty
\big(\int_{x\in \Omega}\psi^{\sigma_{\sym}(a,P),K}(x)\,dx\big)^\frac{1}{2}\\&
\Big(\Vol(\Omega)+\|K\|_{1}^2(\lambda_{\min}(a))^{-1}\int_{x\in \Omega,y\in [0,1]^d} \psi^{a,E}(x,y)\,dx\,dy\Big)^\frac{1}{2}.
\end{split}
\end{equation}
Using Cauchy-Schwartz inequality, lemma \ref{sdjhddhsgddgg7812} and \\$\nabla .\sigma_{\sym}(a,P)\nabla \psi^{\sigma_{\sym}(a,P),K}(x)=-1-\nabla .K(x)\nabla \psi^{\sigma_{\sym}(a,P),K}(x)$, we obtain from equation \eref{daassjhcchdddjkj1s411c6aaassab} that
\begin{equation}\label{daasssjhcchddssadjkj1s411c6aaassab}
\begin{split}
|G_2|\leq &C_{d}\|a+P\|_\infty(\lambda_{\min}(a))^{-\frac{1}{2}}\big(\lambda_{\min}(\sigma_{\sym}(a,P))\big)^{-1}\| \xi\|_\infty
\big(\int_{x\in \Omega,y\in [0,1]^d} \psi^{a,E}(x,y)\,dx\,dy\big)^\frac{1}{2}\\&
\Big(\Vol(\Omega)+\|K\|_{1}^2\big(\lambda_{\min}(\sigma_{\sym}(a,P))\big)^{-1}\int_{x\in \Omega}\psi^{\sigma_{\sym}(a,P),K}(x)\,dx\Big)^\frac{1}{2}.
\end{split}
\end{equation}
Combining \eref{dssjhccssshdddjkj1411c6aaaa}, \eref{dsaysjhcchyydaaadyyydjkj1s411c6aaassab}, \eref{dssjssaashcchdddjkj1411c6ac}, \eref{dssjhcchaadddjkj1411c6aaaabe} and \eref{daasssjhcchddssadjkj1s411c6aaassab} we obtain that
\begin{equation}\label{dsqaysjhcaaydaaadyyaaydjkj1s411c6aaassab}
\begin{split}
\int_{z\in \partial \Omega,y\in [0,1]^d} & \big({^t\nabla} \psi^{a,E}(z,y)a \nabla \psi^{\sigma_{\sym}(a,P),K}(z)\big)  \,dz\,dy \leq C_{d,\|a+P\|_\infty/\lambda_{\min}(a)}(1+ \frac{\|K\|_{1}}{\lambda_{\min}(a)} )\\&\big(\|\xi\|_\infty+\|\nabla \xi\|_\infty\big)
\Big(\int_{x\in \Omega,y\in [0,1]^d} \psi^{a,E}(x,y)\,dx\,dy\int_{x\in \Omega}\psi^{\sigma_{\sym}(a,P),K}(x)\,dx\Big)^\frac{1}{2}
\\&+ C_{d,\|a+P\|_\infty/\lambda_{\min}(a)}\| \xi\|_\infty
\Big(\frac{\Vol(\Omega)}{\lambda_{\min}(a)}\Big)^\frac{1}{2}
\\&\Big(\int_{x\in \Omega}\psi^{\sigma_{\sym}(a,P),K}(x)\,dx+\int_{x\in \Omega,y\in [0,1]^d} \psi^{a,E}(x,y)\,dx\,dy\Big)^\frac{1}{2}.
\end{split}
\end{equation}
Which proves the lemma by optimization on the vector field $\xi$.
\end{proof}

\begin{Proposition}\label{eheheehdudh77122}
We have
\begin{equation}\label{daqasaysjhcchdddjkj1s411c6aaassabew}
\begin{split}
|J|\leq  & R^{-1}C_{d,\|a+P\|_\infty/\lambda_{\min}(a)}(\Theta(\Omega)+1)
 \\&
\Bigg(\big(1+ \frac{\|K\|_{1}}{\lambda_{\min}(a)} \big)
\Big(\int_{x\in \Omega,y\in [0,1]^d} \psi^{a,E}(x,y)\,dx\,dy\int_{x\in \Omega}\psi^{\sigma_{\sym}(a,P),K}(x)\,dx\Big)^\frac{1}{2}
\\&+
\Big(\frac{\Vol(\Omega)}{\lambda_{\min}(a)}\Big)^\frac{1}{2}
\Big(\big(\int_{x\in \Omega}\psi^{\sigma_{\sym}(a,P),K}(x)\,dx\big)^\frac{1}{2}+\big(\int_{x\in \Omega,y\in [0,1]^d} \psi^{a,E}(x,y)\,dx\,dy\big)^\frac{1}{2}\Big)\Bigg).
\end{split}
\end{equation}
\end{Proposition}
\begin{proof}
Using formulae \eref{dssjhcchc6} and \eref{jdhsdhvcjhcvdh891} one obtains that
\begin{equation}\label{dssjhcchdddjkj1411c6}
\begin{split}
J=\sum_{i,j,k=1}^d \int_{x\in \Omega,y\in [0,1]^d} \partial_i \psi^{a,E}(x,y)B_{i,j,k}(x,y) \partial_k \psi^{\sigma_{\sym}(a,P),K}(x)\,dx\,dy
\end{split}
\end{equation}
with
\begin{equation}\label{dssjhcchdddjkj1115c6}
\begin{split}
B_{i,j,k}(x,y)=-\partial_j M^P_{ijk}(Rx+y) +\big(a-P(Rx+y)\big)_{ij}\partial_j \chi^{a,P}_k(Rx+y).
\end{split}
\end{equation}
 Applying formula \eref{sdkddkdjdduu8822} to equation \eref{dssjhcchdddjkj1411c6} first with
$\diiv \xi=\sum_{j=1}^d-\partial_j M^P_{ijk}(Rx+y)$, next with
$\nabla f=\nabla \chi^{a,P}_k(Rx+y)$
we obtain that
\begin{equation}\label{dssjhcchdddjkj1411c6aaaa}
\begin{split}
J=J_1+J_2+J_3
\end{split}
\end{equation}
with (using the skew symmetry of $M^P_{ijk}$ in $ij$)
\begin{equation}\label{dssjhcchdddjkj1411c6aaaab}
\begin{split}
J_1=-R^{-1}\sum_{i,j,k=1}^d \int_{x\in \Omega,y\in [0,1]^d} \partial_j \Big(  a+P(Rx+y)\big)_{ji}\partial_i \psi^{a,E}(x,y)\Big) \chi^{a,P}_k(Rx+y)\partial_k \psi^{\sigma_{\sym}(a,P),K}(x)\,dx\,dy
\end{split}
\end{equation}
\begin{equation}\label{dssjhcchdddjkj1s411c6aaassab}
\begin{split}
J_2=&R^{-1}\sum_{i,j,k=1}^d \int_{x\in \Omega,y\in [0,1]^d}  \partial_i \psi^{a,E}(x,y)\Big( M^P_{ijk}(Rx+y) -\big(a-P(Rx+y)\big)_{ij} \chi^{a,P}_k(Rx+y)\Big) \\&\partial_j\partial_k \psi^{\sigma_{\sym}(a,P),K}(x)\,dx\,dy
\end{split}
\end{equation}
\begin{equation}\label{dsaysjhcchdddjkj1s411c6aaassab}
\begin{split}
J_3=&R^{-1}\sum_{i,j,k=1}^d \int_{z\in \partial \Omega,y\in [0,1]^d}  \partial_i \psi^{a,E}(z,y)\Big( -M^P_{ijk}(Rz+y) +\big(a-P(Rz+y)\big)_{ij} \chi^{a,P}_k(Rz+y)\Big)\\&n_{j}(z)\partial_k \psi^{\sigma_{\sym}(a,P),K}(z) \,\,dz\,dy.
\end{split}
\end{equation}
 Using the fact that $\nabla \psi^{a,E}(z,y)$
and $\nabla \psi^{\sigma_{\sym}(a,P),K}(z)$ are parallel to $n(z)$ at the boundary of $\Omega$ and both heading towards the opposite direction of $n$, we obtain that (using the skew symmetry of $M^P_{ijk}$ and $P_{ij}$ in $ij$)
\begin{equation}\label{dsaysjhcchdddjkj1s411c6aaassabe}
\begin{split}
J_3=&R^{-1} \int_{z\in \partial \Omega,y\in [0,1]^d}  \big({^t\nabla} \psi^{a,E}(z,y)a \nabla \psi^{\sigma_{\sym}(a,P),K}(z)\big)\big(\chi^{a,P}_.(Rz+y).n(z)\big)  \,\,dz\,dy.
\end{split}
\end{equation}
Thus by lemma \ref{sdjdgjhddgg787112} and equation \eref{dkjdjdjj81222231}
\begin{equation}\label{daqasaysjhcchdddjkj1s411c6aaassab}
\begin{split}
|J_3|\leq &R^{-1} \|\chi^{a,P}_.\|_\infty \int_{z\in \partial \Omega,y\in [0,1]^d}  \big({^t\nabla} \psi^{a,E}(z,y)a \nabla \psi^{\sigma_{\sym}(a,P),K}(z)\big) \,dz\,dy
\\\leq & R^{-1}C_{d,\|a+P\|_\infty/\lambda_{\min}(a)}\Theta(\Omega)
 \\&
\Bigg(\big(1+ \frac{\|K\|_{1}}{\lambda_{\min}(a)} \big)
\Big(\int_{x\in \Omega,y\in [0,1]^d} \psi^{a,E}(x,y)\,dx\,dy\int_{x\in \Omega}\psi^{\sigma_{\sym}(a,P),K}(x)\,dx\Big)^\frac{1}{2}
\\&+
\Big(\frac{\Vol(\Omega)}{\lambda_{\min}(a)}\Big)^\frac{1}{2}
\Big(\int_{x\in \Omega}\psi^{\sigma_{\sym}(a,P),K}(x)\,dx+\int_{x\in \Omega,y\in [0,1]^d} \psi^{a,E}(x,y)\,dx\,dy\Big)^\frac{1}{2}\Bigg).
\end{split}
\end{equation}
Using Cauchy-Schwartz inequality and $\nabla .(a+P(Rx+y))\nabla\psi^{a,E}(x,y)=-1-\nabla .K(x)\nabla\psi^{a,E}(x,y)$ we obtain
from equation
\eref{dssjhcchdddjkj1411c6aaaab} and \eref{dkjdjdjj81222231} that
\begin{equation}\label{dssjhccewhaadddjkj1411c6aaaabe}
\begin{split}
|J_1|\leq &C_{d,\|a+P\|_\infty/\lambda_{\min}(a)}R^{-1}
\big(\int_{x\in \Omega}\psi^{\sigma_{\sym}(a,P),K}(x)\,dx\big)^\frac{1}{2}\\&
\Bigg(\Big(\frac{\Vol(\Omega)}{\lambda_{\min}(a)}\Big)^\frac{1}{2}+\frac{\|K\|_{1}}{\lambda_{\min}(a)}\Big(\int_{x\in \Omega,y\in [0,1]^d} \psi^{a,E}(x,y)\,dx\,dy\Big)^\frac{1}{2}\Bigg).
\end{split}
\end{equation}
Using Cauchy-Schwartz inequality, lemma \ref{sdjhddhsgddgg7812} and \\$\nabla .\sigma_{\sym}(a,P)\nabla \psi^{\sigma_{\sym}(a,P),K}(x)=-1-\nabla .K(x)\nabla \psi^{\sigma_{\sym}(a,P),K}(x)$, we obtain from equations \eref{dssjhcchdddjkj1s411c6aaassab}, \eref{dkjdjdjj81222231} and \eref{dbekjbo1} that
\begin{equation}\label{daasssjhsscchewddssadjkj1s411c6aaassab}
\begin{split}
|J_2|\leq &C_{d}R^{-1}C_{d,\|a+P\|_\infty/\lambda_{\min}(a)}
\big(\int_{x\in \Omega,y\in [0,1]^d} \psi^{a,E}(x,y)\,dx\,dy\big)^\frac{1}{2}\\&
\Bigg(\Big(\frac{\Vol(\Omega)}{\lambda_{\min}(a)}\Big)^\frac{1}{2}+\frac{\|K\|_{1}}{\lambda_{\min}(a)}\Big(\int_{x\in \Omega}\psi^{\sigma_{\sym}(a,P),K}(x)\,dx\Big)^\frac{1}{2}\Bigg).
\end{split}
\end{equation}
The proposition \eref{eheheehdudh77122} is then a straightforward combination of \eref{dssjhcchdddjkj1411c6aaaa}, \eref{daqasaysjhcchdddjkj1s411c6aaassab}, \eref{dssjhccewhaadddjkj1411c6aaaabe} and \eref{daasssjhsscchewddssadjkj1s411c6aaassab}.
\end{proof}
We will now need the following lemma whose proof is trivial algebra
\begin{Lemma}\label{ezegzdgdgdzgdg73732}
Assume $X,Y,\delta,\eta>0$ and
\begin{equation}\label{sjdhdbdhdhdbhdhd7712}
(X-Y)^2\leq \delta XY+\eta (X+Y)
\end{equation}
then
\begin{equation}\label{ashshssis7272121}
X^\frac{1}{2}\leq Y^\frac{1}{2} (1+\sqrt{\delta})+\sqrt{\eta}
\end{equation}
and
\begin{equation}\label{ashshssis7272ss121}
X^\frac{1}{2}\geq \big(Y^\frac{1}{2} -\sqrt{\eta}\big) (1+\sqrt{\delta})^{-1}.
\end{equation}
\end{Lemma}
\begin{proof}
The upper root of the equations \eref{sjdhdbdhdhdbhdhd7712} is  $X_0=Y (1+\frac{\delta}{2})+\frac{\eta}{2}+\frac{\sqrt{\Delta}}{2}$  with $\Delta= Y^2 \delta(\delta+4)+ Y \eta (8+2\delta)+\eta^2$. Then by applying Minkowski inequality to $\sqrt{\Delta}$ we obtain
\begin{equation}
X\leq Y (1+\sqrt{\delta})^2 + 2\sqrt{Y}\sqrt{\eta}(1+\sqrt{\delta})+\eta
\end{equation}
which leads to \eref{ashshssis7272121}. The equation \eref{ashshssis7272ss121} is then obtained by the symmetry of \eref{sjdhdbdhdhdbhdhd7712} in $X$ and $Y$.
\end{proof}
Combining proposition \ref{duudgddg7782727717221} and \ref{eheheehdudh77122} with lemma \ref{ezegzdgdgdzgdg73732} we obtain  theorem \ref{dddhdduhdhdhh88221}
\begin{Theorem}\label{dddhdduhdhdhh88221}
There exists a finite function $h:\, (\R^+)^2\rightarrow \R^+$ increasing in each argument such that
the following inequalities are valid
\begin{equation}
X\leq Y (1+\delta)+\eta \quad \text{and}\quad X\geq \big(Y -\eta\big) (1+\delta)^{-1}
\end{equation}
with
\begin{equation}
X:=\Big(\frac{1}{\Vol(\Omega)}\int_{x\in \Omega,y\in [0,1]^d} \psi^{a,E}(x,y)\,dx\,dy\Big)^\frac{1}{4}
\end{equation}
\begin{equation}
Y:=\Big(\frac{1}{\Vol(\Omega)}\int_{x\in \Omega}\psi^{\sigma_{\sym}(a,P),K}(x)\,dx\Big)^\frac{1}{4}
\end{equation}
\begin{equation}
\delta:=R^{-\frac{1}{2}}h\big(d,\frac{1+\|a+P\|_\infty}{\lambda_{\min}(a)}\big)(\Theta(\Omega)+1)^\frac{1}{2}
\big(1+ \|K\|_{1} \big)^\frac{1}{2}
\end{equation}
and
\begin{equation}
\eta:=R^{-\frac{1}{2}}h\big(d,\frac{1+\|a+P\|_\infty}{\lambda_{\min}(a)}\big)(\Theta(\Omega)+1)^\frac{1}{2}.
\end{equation}
\end{Theorem}
\subsubsection{Effect of relative translation on averaging}
For $\Omega$ a bounded open subset of $\R^d$ with smooth boundary and $E$ a skew symmetric matrix with smooth coefficients in $L^\infty_{loc}(\R^d)$ and $a\in \M_{d,\sym}$, let $\psi^{a,E}_{\Omega}(x)$ the solution of $\nabla.(a+E)\nabla \psi^{a,E}_{\Omega}=-1$ in $\Omega$. For $y\in [0,1]^d$ let us introduce the operator $\theta_y$ such that for any function $f$ on $\R^d$, $\theta_y f(x)=f(x+y)$. Using the notation \eref{erjbjdfbjfbfbq}, let us observe that for $y\in [0,1]^d$
\begin{equation}
\psi_\Omega^{a,\theta_{y/R}(S_R P)+K}(x)=\psi_\Omega(x,y).
\end{equation}
\begin{Lemma}\label{dslkdnlkcvnnf91112qqq1}
For  $y\in \R^d$ one has
\begin{equation}\label{HoCoeDNo1zfgscv2868mo1001}
\begin{split}
\int_{x\in \Omega}|\nabla \psi_\Omega^{a,\theta_{y/R}(S_R P+K)}(x)-\nabla \psi_\Omega(x,y)|_{a}^2\leq  C_d (\frac{\|K\|_1}{R \lambda_{\min}(a)})^2 \int_{x\in \Omega} \psi_\Omega^{a,E}(x,y)\,dx.
 \end{split}
\end{equation}
\end{Lemma}
\begin{proof}
Observe that
\begin{equation}
\begin{split}
\nabla .(a+E(x+y/R))&\nabla\big( \psi_\Omega^{a,\theta_{y/R}(S_R P+K)}(x)- \psi_\Omega^{a,E}(x,y)\big)\\&=\nabla .\Big(\big(K(x+y/R)-K(x)\big)\nabla \psi_\Omega^{a,E}(x,y)\Big).
\end{split}
\end{equation}
It follows that
\begin{equation}\label{HoCoeDNo1ffzfgscv2868moo1}
\begin{split}
\int_{x\in \Omega}|\nabla \psi_\Omega^{a,\theta_{y/R}(S_R P+K)}(x)-\nabla \psi_\Omega^{a,E}(x,y)|_{a}^2=&
\int_{x\in \Omega}\big(\nabla \psi_\Omega^{a,\theta_{y/R}(S_R P+K)}(x)-\nabla \psi_\Omega^{a,E}(x,y)\big)\\&\big(K(x+y/R)-K(x)\big)\nabla \psi_\Omega^{a,E}(x,y)
\end{split}
\end{equation}
thus by Cauchy-Schwartz inequality
\begin{equation}\label{HoCoeDNo1ffzfgscv2868moo1q}
\begin{split}
\int_{x\in \Omega}|\nabla \psi_\Omega^{a,\theta_{y/R}(S_R P+K)}(x)-\nabla \psi_\Omega^{a,E}(x,y)|_{a}^2\leq \int_{x\in \Omega}\Big|(K(x+y/R)-K(x))\nabla \psi_\Omega^{a,E}(x,y)\Big|^2_{a^{-1}}
\end{split}
\end{equation}
and the equation \ref{HoCoeDNo1zfgscv2868mo1001} follows easily.
\end{proof}
Now we will need the following lemma
\begin{Lemma}\label{jsdhhdddhhdduudh898121}
For $y\in [0,1]^d$
\begin{equation}\label{ejhebedhdbd771212}
\int_{x\in \Omega}\psi_\Omega^{a,\theta_{y/R}(S_R P+K)}(x)\,dx\leq \int_{x\in \Omega} \psi_\Omega(x,y)\,dx \big(1+ C_d \frac{\|K\|_1}{R \lambda_{\min}(a)}\big)^2
\end{equation}
and
\begin{equation}\label{ejhebedhdbd7712122w}
\int_{x\in \Omega}\psi_\Omega^{a,\theta_{y/R}(S_R P+K)}(x)\,dx\geq \int_{x\in \Omega} \psi_\Omega(x,y)\,dx \big(1+ C_d \frac{\|K\|_1}{R \lambda_{\min}(a)}\big)^{-2}.
\end{equation}
\end{Lemma}
\begin{proof}
Combining the identity
\begin{equation}
\int_{x\in \Omega}\psi_\Omega^{a,\theta_{y/R}(S_R P+K)}(x)\,dx=\int_{x\in \Omega}|\nabla\psi_\Omega^{a,\theta_{y/R}(S_R P+K)}(x)|^2_a\,dx
\end{equation}
with Minkowski inequality we obtain that
\begin{equation}
\begin{split}
\Big(\int_{x\in \Omega}\psi_\Omega^{a,\theta_{y/R}(S_R P+K)}(x)\,dx\Big)^\frac{1}{2}\leq& \Big(\int_{x\in \Omega}|\nabla\psi_\Omega^{a,E}(x,y)|^2_a\,dx\Big)^\frac{1}{2}\\&+\Big(\int_{x\in \Omega}|\nabla \psi_\Omega^{a,\theta_{y/R}(S_R P+K)}(x)-\nabla \psi_\Omega(x,y)|_{a}^2\Big)^\frac{1}{2}
\end{split}
\end{equation}
and the equation \eref{ejhebedhdbd771212} follows from lemma \ref{dslkdnlkcvnnf91112qqq1}. The proof of inequality \eref{ejhebedhdbd7712122w} is similar.
\end{proof}
 We write
\begin{equation}
X(\Omega,a,E):=\big(\Vol(\Omega)\big)^{-1}\int_{x\in \Omega}\psi^{a,E}_\Omega(x)\,dx.
\end{equation}
For $y\in \R^d$ we write $\theta_y \Omega:=\{x+y\,:\, x\in \Omega\}$. From lemma \ref{jsdhhdddhhdduudh898121} we obtain the following proposition
\begin{Proposition}\label{dsaqdjaahydgddgdg777112}
\begin{equation}\label{ejhebedhdbdy771s212e}
X(\theta_{\frac{y}{R}} \Omega,a,S_R P+K) \leq  X( \Omega,a,\theta_{\frac{y}{R}}(S_R P)+K) \big(1+ C_d \frac{\|K\|_1}{R \lambda_{\min}(a)}\big)^2
\end{equation}
and
\begin{equation}\label{ejhebyedhdbd7712122we}
 X(\theta_{\frac{y}{R}} \Omega,a,S_R P+K)\geq X( \Omega,a,\theta_{\frac{y}{R}}(S_R P)+K) \big(1+ C_d \frac{\|K\|_1}{R \lambda_{\min}(a)}\big)^{-2}.
\end{equation}
\end{Proposition}

\subsubsection{Reverse iteration to obtain supper-diffusion}
It is easy to obtain from theorem \eref{xsddhdxhxvhxv898812}, that for any $\gamma >0$,
\begin{equation}\label{dddjddkjsjshsjha}
X(\Omega,a,\gamma E)= \gamma^{-1}X(\Omega,\gamma^{-1}a,E).
\end{equation}
Moreover for $R>0$, writing $S_R \Omega:=\{x\in \R^d\,:\, R^{-1}x\in \Omega\}$ it is easy to obtain by scaling that
\begin{equation}\label{dddjddkjsjshsjha1}
X(S_R \Omega,a,  E)=R^2 X(\Omega,a, S_R E).
\end{equation}
Let us write for $0\leq p\leq n-1$,
\begin{equation}\label{dshdhddhkdsjhd771221}
Z(p,B):=\Big(\int_{(y_p,\ldots,y_{n-1})\in [0,1]^{d\times (n-p)}}X\big( S_{R_n}\prod_{k=p}^{n-1}\theta_{\frac{y_k R_k}{R_n}}\Omega,B,\frac{1}{\gamma_p}\Gamma^{p,\infty}\big)\,dy_p\ldots\,dy_{n-1}\Big)^\frac{1}{4}.
\end{equation}
We will need the following proposition
\begin{Proposition}\label{sdsjhhdhdvv78727281}
There exists a finite increasing function $F: (R^+)^2\rightarrow \R^+$ such that for
\begin{equation}\label{geexkjddheeuzdhdzeezh8sa761}
\rho_{\min}>\gamma_{\max} Q_n \big(1+\frac{R_n}{\big(Z(n,A^{n})\big)^2}\big)
\end{equation}
one has
\begin{equation}\label{shyqgsfgsxfdadhddajhdgd7887712aaqq}
\begin{split}
Z(0,A_0)\geq  0.5\big(\frac{R_n^2}{\gamma_{n}}\big)^\frac{1}{4}\frac{Z\big(n,A^{n}\big)}{R_n^\frac{1}{2}} \big(1+\big(\frac{Q_n \gamma_{\max}}{\rho_{\min}}\big)^{\frac{1}{2}}\big)^{-n}
\end{split}
\end{equation}
and
\begin{equation}\label{shyqxgsfgsxfdadhddajhdgd7887712aaqq}
\begin{split}
Z(0,A_0)\leq 2 \big(\frac{R_n^2}{\gamma_{n}}\big)^\frac{1}{4}\frac{Z\big(n,A^{n}\big)}{R_n^\frac{1}{2}} \big(1+\big(\frac{Q_n \gamma_{\max}}{\rho_{\min}}\big)^{\frac{1}{2}}\big)^{n}
\end{split}
\end{equation}
with
\begin{equation}
Q_n:=F\big(d,\frac{1+K_0}{\lambda^-_n}+\mu_n\big)
(\Theta(\Omega)+1) (1+K_{1})
\end{equation}
where $\lambda^-_n$ is the stability of the \emph{renormalization core} $(A^n)_{n\in \N}$ and $\mu_n$ its anisotropic distortion.
\end{Proposition}
\begin{proof}
From proposition \ref{dsaqdjaahydgddgdg777112} and equation \eref{dddjddkjsjshsjha1} we obtain that
\begin{equation}\label{dshdhddhkdsjhd771221aayx}
\begin{split}
Z(p,B)\leq &\Big(\int_{(y_p,\ldots,y_{n-1})\in [0,1]^{d\times (n-p)}}X\big( \prod_{k=p+1}^{n-1}\theta_{\frac{y_k R_k}{R_n}}\Omega,B,S_{\frac{R_n}{R_p}}\theta_{y_p}E^p+\frac{1}{\gamma_p}S_{R_n}\Gamma^{p+1,\infty}\big)\,dy_p\ldots\,dy_{n-1}\Big)^\frac{1}{4}\\& R_n^\frac{1}{2}\big(1+ C_d \frac{\|\frac{1}{\gamma_p}S_{R_n}\Gamma^{p+1,\infty}\|_1 R_p}{R_n \lambda_{\min}(B)}\big)^{\frac{1}{2}}
\end{split}
\end{equation}
and
\begin{equation}\label{dshdhddhkdsjhd771221aa2yx}
\begin{split}
Z(p,B)\geq &\Big(\int_{(y_p,\ldots,y_{n-1})\in [0,1]^{d\times (n-p)}}X\big( \prod_{k=p+1}^{n-1}\theta_{\frac{y_k R_k}{R_n}}\Omega,B,S_{\frac{R_n}{R_p}}\theta_{y_p}E^p+\frac{1}{\gamma_p}S_{R_n}\Gamma^{p+1,\infty}\big)\,dy_p\ldots\,dy_{n-1}\Big)^\frac{1}{4}\\&R_n^\frac{1}{2}\big(1+ C_d \frac{\|\frac{1}{\gamma_p}S_{R_n}\Gamma^{p+1,\infty}\|_1 R_p}{R_n \lambda_{\min}(B)}\big)^{-\frac{1}{2}}.
\end{split}
\end{equation}
Now let us observe that
\begin{equation}\label{djshdgsgdgdgd66121}
\begin{split}
\|S_{R_n} \frac{1}{\gamma_p}\Gamma^{p+1,\infty}\|_1&\leq K_1 \frac{R_n}{R_{p+1}} \sum_{k=p+1}^\infty (\gamma_k/\gamma_p) (R_{p+1}/ R_{k})
\\\leq & K_1 \frac{R_n}{R_{p+1}} \frac{\gamma_{p+1}}{\gamma_p} (1-\gamma_{\max}/\rho_{\min})^{-1}.
\end{split}
\end{equation}
Combining \eref{dshdhddhkdsjhd771221aayx} and \eref{dshdhddhkdsjhd771221aa2yx} with  theorem \ref{dddhdduhdhdhh88221} (with $R=R_n/R_p$, $P=E^p$, $K=S_{R_n}\Gamma^{p+1,\infty}/\gamma_p$) and \eref{dddjddkjsjshsjha} one obtains that
\begin{equation}\label{ddddddndjun8811221ae}
\begin{split}
Z(p,B)\leq \big(\frac{\gamma_{p}}{\gamma_{p+1}}\big)^\frac{1}{4}Z\big(p+1,\frac{\gamma_p}{\gamma_{p+1}}\sigma_{sym}(B,E^p)\big) \big(1+\delta_{p}(B)\big)+\eta_p(B)
\end{split}
\end{equation}
and
\begin{equation}\label{dddqwdddndjun8811221ae}
\begin{split}
Z(p,B)\geq \Big(\big(\frac{\gamma_{p}}{\gamma_{p+1}}\big)^\frac{1}{4}Z\big(p+1,\frac{\gamma_p}{\gamma_{p+1}}\sigma_{sym}(B,E^p)\big)
-\eta_p(B)\Big) \big(1+\delta_{p}(B)\big)^{-1}
\end{split}
\end{equation}
with
\begin{equation}\label{dskjsddddh7712}
\begin{split}
\delta_p(B):=\big(\frac{R_p}{R_{p+1}}\frac{\gamma_{p+1}}{\gamma_p}\big)^{\frac{1}{2}}h\big(d,\frac{1+\lambda_{\max}(B)+K_0}{\lambda_{\min}(B)}\big)
(\Theta(\Omega)+1)^\frac{1}{2} (1+K_{1})^\frac{1}{2} (1-\gamma_{\max}/\rho_{\min})^{-1}
\end{split}
\end{equation}
and
\begin{equation}\label{sdjdhhghdgddgdhgdgd721}
\eta_p(B):=R_p^\frac{1}{2}h\big(d,\frac{1+\lambda_{\max}(B)+K_0}{\lambda_{\min}(B)}\big)(\Theta(\Omega)+1)^\frac{1}{2} (1-\gamma_{\max}/\rho_{\min})^{-\frac{1}{2}}.
\end{equation}
Where, in \eref{dskjsddddh7712}  we have used the  inequality \eref{djshdgsgdgdgd66121} and we have integrated the error terms
 involving $B$ appearing in \eref{dshdhddhkdsjhd771221aayx} and \eref{dshdhddhkdsjhd771221aa2yx} in
 the function $h$ and used the assumption
$K_1 \gamma_{\max} \leq \rho_{\min}$. Then one obtains from \eref{ddddddndjun8811221ae} by a simple induction
that for $n\geq 2$
\begin{equation}\label{shgsfgsfddhddjhdgd7887712}
\begin{split}
Z(0,A_0)\leq & \big(\frac{\gamma_{0}}{\gamma_{n}}\big)^\frac{1}{4}Z\big(n,A^{n}\big)\prod_{k=0}^{n-1} \big(1+\delta_k(A_k)\big)
\\&+ \sum_{p=0}^{n-2}\eta_{p+1}(A^{p+1}) \big(\frac{\gamma_{0}}{\gamma_{p+1}}\big)^\frac{1}{4}\prod_{k=0}^p \big(1+\delta_k(A_k)\big)
\\&+\eta_0(A_0).
\end{split}
\end{equation}
Similarly one obtains from \eref{dddqwdddndjun8811221ae} by a simple induction that for $n\geq 2$
\begin{equation}\label{hdsdhgdhdgdgzsdgdzdsz8771221}
\begin{split}
Z(0,A_0)\geq & \big(\frac{\gamma_{0}}{\gamma_{n}}\big)^\frac{1}{4}Z\big(n,A^{n}\big)\prod_{k=0}^{n-1} \big(1+\delta_k(A_k)\big)^{-1}
\\&- \sum_{p=0}^{n-2}\eta_{p+1}(A^{p+1}) \big(\frac{\gamma_{0}}{\gamma_{p+1}}\big)^\frac{1}{4}\prod_{k=0}^p \big(1+\delta_k(A_k)\big)^{-1}
\\&-\eta_0(A_0)\big(1+\delta_0(A_0)\big)^{-1}
\end{split}
\end{equation}
Where $\big(A^{k}\big)_{k\in \N}$, is the \emph{renormalization core} \eref{wdkbkwjdebkljb81}.
Now combining \eref{hdsdhgdhdgdgzsdgdzdsz8771221} and \eref{dddjddkjsjshsjha1} we obtain that
\begin{equation}\label{shgqasfgsfddhddjhdgd7887712aaqq}
\begin{split}
Z(0,A_0)\geq & \big(\frac{R_n^2}{\gamma_{n}}\big)^\frac{1}{4}\frac{Z\big(n,A^{n}\big)}{R_n^\frac{1}{2}}
\prod_{k=0}^{n-1} \big(1+\big(\frac{\gamma_{k+1}}{r_{k+1}\gamma_k}\big)^{\frac{1}{2}}\delta(A_k)\big)^{-1}
\\&-\sum_{p=0}^{n-2}R_{p+1}^\frac{1}{2}\delta(A^{p+1}) \big(\frac{\gamma_{0}}{\gamma_{p+1}}\big)^\frac{1}{4}\prod_{k=0}^p \big(1+\big(\frac{\gamma_{k+1}}{r_{k+1}\gamma_k}\big)^{\frac{1}{2}}\delta(A_k)\big)^{-1}
-\delta(A_0)\big(1+\delta(A_0)\big)^{-1}
\end{split}
\end{equation}
with
\begin{equation}\label{dskjsqwddddh7712}
\begin{split}
\delta(B):=h\big(d,\frac{1+\lambda_{\max}(B)+K_0}{\lambda_{\min}(B)}\big)
(\Theta(\Omega)+1)^\frac{1}{2} (1+K_{1})^\frac{1}{2} (1-\gamma_{\max}/\rho_{\min})^{-1}.
\end{split}
\end{equation}
Thus
\begin{equation}\label{shyqgsfgsfdadhddajhdgd7887712aaqq}
\begin{split}
Z(0,A_0)\geq  (1-\zeta_i)\big(\frac{R_n^2}{\gamma_{n}}\big)^\frac{1}{4}\frac{Z\big(n,A^{n}\big)}{R_n^\frac{1}{2}}\prod_{k=0}^{n-1} \big(1+\big(\frac{\gamma_{k+1}}{r_{k+1}\gamma_k}\big)^{\frac{1}{2}}\delta(A_k)\big)^{-1}
\end{split}
\end{equation}
with
\begin{equation}\label{shgsfgsfddhddaqjhdagd7887712aaqq}
\begin{split}
\zeta_i \leq &  \frac{R_n^\frac{1}{2}}{Z\big(n,A^{n}\big)} \Bigg(
\sum_{p=0}^{n-2}\big(\frac{R_{p+1}^2 \gamma_n}{R_n^2 \gamma_{p+1}}\big)^\frac{1}{4}\delta(A^{p+1}) \prod_{k=p+1}^{n-1} \big(1+\big(\frac{\gamma_{k+1}}{r_{k+1}\gamma_k}\big)^{\frac{1}{2}}\delta(A_k)\big)
\\&+\big(\frac{ \gamma_n}{R_n^2}\big)^\frac{1}{4}\prod_{k=0}^{n-1} \big(1+\big(\frac{\gamma_{k+1}}{r_{k+1}\gamma_k}\big)^{\frac{1}{2}}\delta(A_k)\big)\Bigg).
\end{split}
\end{equation}
Thus writing
\begin{equation}\label{ewweeheuheueh712}
W_n:=h\big(d,\frac{1+K_0}{\lambda^-_n}+\mu_n\big)
(\Theta(\Omega)+1)^\frac{1}{2} (1+K_{1})^\frac{1}{2} (1-\gamma_{\max}/\rho_{\min})^{-1}
\end{equation}
we obtain from \eref{shgsfgsfddhddaqjhdagd7887712aaqq} under the assumption $\rho_{\min}^2>\gamma_{\max} (16+W_n^2)$ that
\begin{equation}\label{shgsfssgsfddhddaqjhdagd7887712aaqq}
\begin{split}
\zeta_i \leq &  \frac{R_n^\frac{1}{2}}{Z\big(n,A^{n}\big)}4W_n \big(\frac{\gamma_{\max}}{\rho_{\min}^2}\big)^\frac{1}{4}.
\end{split}
\end{equation}
Thus for
\begin{equation}\label{geekjddheeuzdhdzeezh8761}
\rho_{\min}>\gamma_{\max}^\frac{1}{2}64 \big(1+W_n (1+\frac{R_n^\frac{1}{2}}{Z\big(n,A^{n}\big)})\big)^2
\end{equation}
$\zeta_i<0.5$ and $\zeta_i$ acts as an error term in the inequality \eref{shyqgsfgsfdadhddajhdgd7887712aaqq}.
Then combining \eref{shyqgsfgsfdadhddajhdgd7887712aaqq}, \eref{geekjddheeuzdhdzeezh8761} and
\eref{ewweeheuheueh712} we obtain the control \eref{shyqgsfgsxfdadhddajhdgd7887712aaqq}. Moreover, we obtain
from \eref{dddjddkjsjshsjha1} and \eref{shgsfgsfddhddjhdgd7887712}  that
\begin{equation}\label{shgsfgsfddhddjhdgd7887712aaqq}
\begin{split}
Z(0,A_0)\leq & \big(R_n^2\frac{\gamma_{0}}{\gamma_{n}}\big)^\frac{1}{4}\frac{Z\big(n,A^{n}\big)}{R_n^\frac{1}{2}}\prod_{k=0}^{n-1} \big(1+\big(\frac{\gamma_{k+1}}{r_{k+1}\gamma_k}\big)^{\frac{1}{2}}\delta(A_k)\big)
\\&+ \sum_{p=0}^{n-2}R_{p+1}^\frac{1}{2}\delta(A^{p+1}) \big(\frac{\gamma_{0}}{\gamma_{p+1}}\big)^\frac{1}{4}\prod_{k=0}^p \big(1+\big(\frac{\gamma_{k+1}}{r_{k+1}\gamma_k}\big)^{\frac{1}{2}}\delta(A_k)\big)
+\delta(A_0).
\end{split}
\end{equation}
From this point the proof of equation \eref{shyqxgsfgsxfdadhddajhdgd7887712aaqq} is similar to the one of equation \eref{shyqgsfgsxfdadhddajhdgd7887712aaqq}
\end{proof}
We will need the following lemma
\begin{Lemma}
We have
\begin{equation}\label{dsjhddgdhgd87791211}
Z(n,A^{n})\leq R_n^\frac{1}{2}\big(X(\Omega, A_n,0)\big)^\frac{1}{4}
\end{equation}
and
\begin{equation}\label{sjhdgddhgdgddzgddz8772612}
\begin{split}
Z(n,A^{n})\geq R_n^\frac{1}{2} \big(X(\Omega, I_d,0)\big)^\frac{1}{4}\big(\lambda_{\max}(A^{n})\big)^{-\frac{1}{4}}
\Big(1+\gamma_{n}^{-2}\sup_{x\in S_{R_n}\Omega}\lambda_{\max}\big({^t\Gamma^{n,\infty}(x)\Gamma^{n,\infty}(x)}\big)\Big)^{-\frac{1}{4}}.
\end{split}
\end{equation}
\end{Lemma}
\begin{proof}
Equations \eref{dsjhddgdhgd87791211} and \eref{sjhdgddhgdgddzgddz8772612} are an easy application of theorem \ref{xsddhdxhxvhxv898812}, corollary \ref{xsddhdxhxvhxv883398812} and equation \eref{dddjddkjsjshsjha1}.
\end{proof}
For $r>0$ we write $T[r] \Omega$ the set of $x\in\R^d$ such that there exists $y\in \Omega$ with $|x-y|\leq r\sqrt{d} $. We write $T[-r] \Omega$ the set of $x\in\Omega$ such that there exists $y\not \in \Omega$ with $|x-y|> r\sqrt{d}$.
From equation \eref{dshdhddhkdsjhd771221}, using $\prod_{k=0}^{n-1}\theta_{\frac{y_k R_k}{R_n}}\Omega \subset T[(\rho_{\min}-1)^{-1}]\Omega$ we obtain that
\begin{equation}\label{dsayhdhdaadhkdsjhda771221}
X\big( S_{R_n}T[(\rho_{\min}-1)^{-1}]\Omega,A_0,\Gamma^{0,\infty}\big)\geq \big(Z(0,A_0)\big)^\frac{1}{4}.
\end{equation}
Similarly we obtain that
\begin{equation}\label{dsayhdhdaadhkdasjhda771221}
X\big( S_{R_n}T[-(\rho_{\min}-1)^{-1}]\Omega,A_0,\Gamma^{0,\infty}\big)\leq \big(Z(0,A_0)\big)^\frac{1}{4}.
\end{equation}
We will need the following proposition
\begin{Proposition}\label{hasgddfdgfdd7877761221}
There exists a finite increasing function $F: (R^+)^2\rightarrow \R^+$ such that for
\begin{equation}\label{geexkjddheeuzdhdzeezh8sa761a}
\rho_{\min}>\gamma_{\max} Q_n \quad \text{and} \quad r >R_1
\end{equation}
one has
\begin{equation}\label{shyqgsfgsxfdadhddajhdgd7887712aaqqa}
\begin{split}
X\big(B(0,r),A_0,\Gamma^{0,\infty}\big)\geq  \frac{C_{d,K_0}}{\gamma_{\max}^2}\frac{r^2}{\gamma_{n}\lambda_{\max}(A_n)} \big(1+\big(\frac{Q_n \gamma_{\max}}{\rho_{\min}}\big)^{\frac{1}{2}}\big)^{-4n}
\end{split}
\end{equation}
and
\begin{equation}\label{shyqxgsfgsxfdaaydhddajhdgd7887712aaqqa}
\begin{split}
X\big(B(0,r),A_0,\Gamma^{0,\infty}\big)\leq  \frac{r^2}{\gamma_{n}\lambda_{\max}(A_n)} \big(1+\big(\frac{Q_n \gamma_{\max}}{\rho_{\min}}\big)^{\frac{1}{2}}\big)^{4n}
\end{split}
\end{equation}
with
\begin{equation}
n=\sup\{p\in \N\,:\, R_p\leq r\}
\end{equation}
\begin{equation}
Q_n:=F\big(d,\frac{1+K_0}{\lambda^-_n}+\mu_n\big)
 (1+K_{1})
\end{equation}
where $\lambda^-_n$ is the stability of the \emph{renormalization core} $(A_n)_{n\in \N}$ and $\mu_n$ its anisotropic distortion.
\end{Proposition}
\begin{proof}
Taking $\Omega:=B\big(0,(r/R_n)-\sqrt{d} (\rho_{\min}-1)^{-1}\big)$ in  equation \eref{dsayhdhdaadhkdsjhda771221} we obtain from equation \eref{shyqgsfgsxfdadhddajhdgd7887712aaqq} of proposition \ref{sdsjhhdhdvv78727281}  and \eref{sjhdgddhgdgddzgddz8772612} that
\begin{equation}\label{shyqgsfgsxfdadhddajhdgd7887712aaqbbqa}
\begin{split}
X\big(B(0,r),A_0,\Gamma^{0,\infty}\big)\geq&  (0.5)^4 \frac{R_n^2}{\gamma_{n}\lambda_{\max}(A_n)}X\big(B(0,r/R_n), I_d,0\big)
 \big(1+\big(\frac{Q_n \gamma_{\max}}{\rho_{\min}}\big)^{\frac{1}{2}}\big)^{-4n}
\\&\Big(1+\gamma_{n}^{-2}\sup_{x\in B(0,r)}\lambda_{\max}\big({^t\Gamma^{n,\infty}(x)\Gamma^{n,\infty}(x)}\big)\Big)^{-1}.
\end{split}
\end{equation}
Which leads to \eref{shyqgsfgsxfdadhddajhdgd7887712aaqqa} by \eref{geexkjddheeuzdhdzeezh8sa761a}, incorporating the new constants in $Q_n$ (observing that for $r\geq 1$, $\Theta(B(0,r))$ is uniformly bounded away from infinity by a constant depending only on the dimension) and using
\begin{equation}
\begin{split}
\sup_{x\in B(0,r)}\gamma_{n}^{-2} \lambda_{\max}\big({^t\Gamma^{n,\infty}(x)\Gamma^{n,\infty}(x)}\big)\leq C_d \Big(K_0 +\gamma_{\max} (1-\frac{\gamma_{\max}}{\rho_{\min}})^{-1} \frac{r}{R_{n+1}}\Big)^2.
\end{split}
\end{equation}
The proof of \eref{shyqxgsfgsxfdaaydhddajhdgd7887712aaqqa} follows similarly by taking $\Omega:=B\big(0,(r/R_n)+\sqrt{d} (\rho_{\min}-1)^{-1}\big)$ in  equation \eref{dsayhdhdaadhkdasjhda771221} and using equation \eref{shyqxgsfgsxfdadhddajhdgd7887712aaqq} of proposition \ref{sdsjhhdhdvv78727281}  and \eref{dsjhddgdhgd87791211}.
\end{proof}
Let us write
\begin{equation}
n(r):=\sup\{p\in \N\,:\, R_p\leq r\}.
\end{equation}
From proposition \eref{hasgddfdgfdd7877761221} we easily deduce the following theorem
\begin{Theorem}\label{hasgddfdgfwdsdxx7877761221y}
There exists a finite increasing function $F: (R^+)^2\rightarrow \R^+$ and a function $|C(r)|\leq
C(d,K_0,\gamma_{\max})$
 such that for
\begin{equation}\label{geexkjddwheesuzdhdzeezh8sa761ay}
\frac{\rho_{\min}}{\gamma_{\max}}>Q(r)  \quad\text{and}\quad r>R_1
\end{equation}
one has
\begin{equation}\label{shyqgsfgsxfdadhddaajhdgd7887712aaqqay}
\begin{split}
\frac{1}{\Vol\big(B(0,r)\big)}\int_{B(0,r)}\E_x\big[\tau(r)\big]\,dx=\frac{r^{2-\nu(r)}}{\lambda_{\max}(A^{n(r)})}
\end{split}
\end{equation}
with
\begin{equation}\label{skdhdgddzwzuzdgz712}
\nu(r)=\frac{\ln \gamma_{n(r)}}{\ln r}\big(1+\epsilon(r)\big)+\frac{C(r)}{\ln r}.
\end{equation}
and
\begin{equation}
|\epsilon(r)| \leq \big(\frac{Q(r) \gamma_{\max}}{\rho_{\min}}\big)^{\frac{1}{2}}
\end{equation}
with
\begin{equation}\label{sdhdgdfd652gs62}
Q(r):=\frac{1}{(\ln \gamma_{\min})^2}F\big(d,\frac{1+K_0}{\lambda^-_{n(r)}}+\mu_{n(r)}\big) (1+K_1)
\end{equation}
\end{Theorem}
Then theorem \ref{hasgddfdgfwdysdxx7877761221y} is a simplified version of theorem
\ref{hasgddfdgfwdsdxx7877761221y} (using theorem \ref{djshddaqadsdbhdb881}). Now we will show that the anomalous
fast behavior of the exit times from $B(0,r)$ is a super-diffusive phenomenon and not a convective phenomenon.
We will consider $\E_{m_{r,l}}\big[\tau(r,l)\big]$ defined by equation
\eref{shyqgsfgsxfaqydadhddaajhdgd788771s2qaqaaqqay}. The following theorem implies theorem
\ref{hasgddfdgfwasdsd7q877761221by}.
\begin{Theorem}\label{hasgddfdgfwdsd7877761221by}
There exists a finite increasing function $F: (R^+)^2\rightarrow \R^+$ such that for
\begin{equation}\label{geexkjddwheesuzdhdzeezh8sa761aby}
\rho_{\min}>\gamma_{\max} 10 Q(r) \quad\text{and}\quad r>R_1
\end{equation}
one has
\begin{equation}\label{shyqgsfgsxfdadhddaajhdgd7887712aaqqaby}
\begin{split}
\lim_{l\rightarrow \infty} \Big(\Vol\big(B(0,r,l)\big)\Big)^{-1}\int_{(y,z)\in B(0,r,l)}\E_{y,z}\Big[\tau\big(B(0,r,l)\big)\Big]=\frac{r^{2-\nu(r)}}{\lambda_{\max}(A^{n(r)})}
\end{split}
\end{equation}
where $\nu(r)$ is given by \eref{skdhdgddzwzuzdgz712} and $Q(r)$ by \eref{sdhdgdfd652gs62}.
\end{Theorem}
\begin{proof}
Let us observe that
\begin{equation}
{B}(0,r,l)\subset \hat{B}(0,r,l)\quad \text{and}\quad \hat{B}(0,r,l) \subset {B}(0,r,l+r).
\end{equation}
Thus, it is sufficient to control exit times from $\hat{B}(0,r,l)$ in order to prove theorem \ref{hasgddfdgfwdsd7877761221by}. Now let us observe that the diffusion $(y_t,z_t)$ is associated to the following generator $L$ acting on  $f\in C^\infty_0(\R^d \times \R^d$).
\begin{equation}
L f(y,z):= \nabla_y.(\kappa I_d + \Gamma(y)-\Gamma(z)) \nabla_y+\nabla_z .(\kappa I_d + \Gamma(z)-\Gamma(y)) \nabla_z.
\end{equation}
Thus one can apply proposition \ref{sdsjhhdhdvv78727281} with $\Omega=\hat{B}(0,r,l)$. Let us
observe that the \emph{renormalization core} associated to $(y_t,z_t)$ is
\begin{equation}
\begin{pmatrix}
A_k & 0 \\
0 & A_k
\end{pmatrix}
\end{equation}
 Moreover it is easy to observe that $\Theta \big(\hat{B}(0,r,l)\big)$ is bounded uniformly away from infinity on $r\leq l$ and that
\begin{equation}
\|\Gamma^{n,\infty}(y)-\Gamma^{n,\infty}(z)\|\leq C_d \sum_{k=n}^\infty C_d K_1\gamma_k R_k^{-1}|y-z|.
\end{equation}
From this point the proof of theorem \ref{hasgddfdgfwdsd7877761221by} is trivially
 similar to the one of theorem \ref{hasgddfdgfwdsdxx7877761221y}.
\end{proof}

\paragraph{Super-diffusion  as a common event}
Let us write $G(r)$ set of points of $B(0,r)$ such that if $y_t$ starts from those points, its exit time from $B(0,r)$ is anomalously fast with probability asymptotically close to one.
We also write $G(r,l)$ set of points of $B(0,r,l)$ such that if $(y_t,z_t)$ starts for those points, their separation time is anomalously fast with probability asymptotically close to one.
More precisely let us write
\begin{equation}
\delta(r)=\frac{\ln \gamma_{n(r)}}{\ln r}\big(1-3C(r)\big)-\frac{C_{d,K_0,\gamma_{\max}}}{\ln r}
\end{equation}
with
\begin{equation}
C(r) = \big(\frac{Q(r) \gamma_{\max}}{\rho_{\min}}\big)^{\frac{1}{2}}.
\end{equation}
Where  $Q(r)$ is given by \eref{sdhdgdfd652gs62}. Let us write
\begin{equation}
\epsilon_2(r) =\exp\big(-\ln \gamma_{n(r)} C(r)\big)
\end{equation}
We will consider
$$
G(r):=\Big\{x\in B(0,r)\,:\,\P_x[\tau(r) \leq \frac{r^{2-\delta(r)}}{\lambda_{\max}(A^{n(r)})}] \geq  1- \epsilon_2(r) \Big\}
$$
 and
$$
G(r,l):=\Big\{(y,z)\in B(0,r,l)\,:\,\P_{y,z}\Big[\tau\big(B(0,r,l)\big) \leq \frac{r^{2-\delta(r)}}{\lambda_{\max}(A^{n(r)})}\Big] \geq  1- \epsilon_2(r) \Big\}
$$
Let us write $m_r, m_{r,l}$ the Lebesgue probability measure defined on $B(0,r)$ and $B(0,r,l)$ by
\begin{equation}
m_r(G(r)):=\frac{\int_{G(r)}dx}{\int_{B(0,r)}dx}
\end{equation}
\begin{equation}
m_{r,l}(G(r,l)):=\frac{\int_{(y,z)\in G(r,l)}\,dy\,dz}{\int_{(y,z)\in B(0,r,l)}\,dy\,dz}
\end{equation}
A trivial consequence of theorem \ref{hasgddfdgfwdsdxx7877761221y} and \ref{hasgddfdgfwdsd7877761221by} is the following theorem.
\begin{Theorem}\label{skdujdh983indn37d}
There exists a finite increasing function $F: (R^+)^2\rightarrow \R^+$ such that for
\begin{equation}\label{geexkjddwheesswuzdhdzeezh8sa761aby}
\rho_{\min}>10 \gamma_{\max} Q(r) \quad\text{and}\quad r>R_1
\end{equation}
\begin{equation}
m_r(G(r))\geq 1-\epsilon_2(r)
\end{equation}
\begin{equation}
m_{r,l}(G(r,l))\geq 1-\epsilon_2(r)
\end{equation}
\end{Theorem}
Theorem \ref{sjdhdjgdd87237sg731} is a particular case of theorem \ref{skdujdh983indn37d}.

\paragraph*{Acknowledgments}
Part of this work was supported by the Aly Kaufman fellowship. The author would like to thank F. Castell, Y.
Velenik and G. Ben Arous for reading the manuscript and A. Majda, T. Hou and P.E. Dimotakis for useful
discussions.

\newcommand{\etalchar}[1]{$^{#1}$}

\end{document}